\documentclass[prl,aps,twocolumn]{revtex4}
\usepackage[dvipdfm]{graphicx}
\usepackage{dcolumn}
\usepackage{bm}
\usepackage{color}
\usepackage{ulem}

\begin{document}
\newcommand{\bR}{\mbox{\boldmath $R$}}
\newcommand{\tr}[1]{\textcolor{black}{#1}}
\newcommand{\trs}[1]{\textcolor{black}{\sout{#1}}}
\newcommand{\tb}[1]{#1}
\newcommand{\tbs}[1]{\textcolor{black}{\sout{#1}}}
\newcommand{\Ha}{\mathcal{H}}
\newcommand{\mh}{\mathsf{h}}
\newcommand{\mA}{\mathsf{A}}
\newcommand{\mB}{\mathsf{B}}
\newcommand{\mC}{\mathsf{C}}
\newcommand{\mS}{\mathsf{S}}
\newcommand{\mU}{\mathsf{U}}
\newcommand{\mX}{\mathsf{X}}
\newcommand{\sP}{\mathcal{P}}
\newcommand{\sL}{\mathcal{L}}
\newcommand{\sO}{\mathcal{O}}
\newcommand{\la}{\langle}
\newcommand{\ra}{\rangle}
\newcommand{\ga}{\alpha}
\newcommand{\gb}{\beta}
\newcommand{\gc}{\gamma}
\newcommand{\gs}{\sigma}
\newcommand{\vk}{{\bm{k}}}
\newcommand{\vq}{{\bm{q}}}
\newcommand{\vR}{{\bm{R}}}
\newcommand{\vQ}{{\bm{Q}}}
\newcommand{\vga}{{\bm{\alpha}}}
\newcommand{\vgc}{{\bm{\gamma}}}
\newcommand{\Ns}{N_{\text{s}}}
\newcommand{\vx}{{\bf r}}
\newcommand{\vG}{{\bf G}}
\newcommand{\avrg}[1]{\left\langle #1 \right\rangle}
\newcommand{\eqsa}[1]{\begin{eqnarray} #1 \end{eqnarray}}
\newcommand{\eqwd}[1]{\begin{widetext}\begin{eqnarray} #1 \end{eqnarray}\end{widetext}}
\newcommand{\hatd}[2]{\hat{ #1 }^{\dagger}_{ #2 }}
\newcommand{\hatn}[2]{\hat{ #1 }^{\ }_{ #2 }}
\newcommand{\wdtd}[2]{\widetilde{ #1 }^{\dagger}_{ #2 }}
\newcommand{\wdtn}[2]{\widetilde{ #1 }^{\ }_{ #2 }}
\newcommand{\cond}[1]{\overline{ #1 }_{0}}
\newcommand{\conp}[2]{\overline{ #1 }_{0#2}}
\newcommand{\nn}{\nonumber\\}
\newcommand{\cdt}{$\cdot$}
\newcommand{\bra}[1]{\langle#1|}
\newcommand{\ket}[1]{|#1\rangle}
\newcommand{\braket}[2]{\langle #1 | #2 \rangle}
\newcommand{\bvec}[1]{\mbox{\boldmath$#1$}}
\newcommand{\blue}[1]{{#1}}
\newcommand{\bl}[1]{{#1}}
\newcommand{\bn}[1]{\textcolor{black}{#1}}
\newcommand{\rr}[1]{{#1}}
\newcommand{\bu}[1]{\textcolor{black}{#1}}
\newcommand{\mgt}[1]{\textcolor{black}{#1}}
\newcommand{\mg}[1]{#1}
\newcommand{\red}[1]{{#1}}
\newcommand{\fj}[1]{{#1}}
\newcommand{\green}[1]{{#1}}
\newcommand{\gr}[1]{\textcolor{black}{#1}}
\definecolor{green}{rgb}{0,0.5,0.1}
\definecolor{blue}{rgb}{0,0,0.8}
\preprint{APS/123-QED}

\title{
First-Principles Study of the Honeycomb-Lattice Iridates Na$_2$IrO$_3$ in
the Presence of Strong Spin-Orbit Interaction and Electron Correlations
}
\author{Youhei Yamaji, 
Yusuke Nomura, Moyuru Kurita, Ryotaro Arita and Masatoshi Imada
}
\affiliation{Department of Applied Physics, University of Tokyo, Hongo, Bunkyo-ku, Tokyo, 113-8656, Japan.}%
\date{\today}

\begin{abstract}
An effective low-energy Hamiltonian of itinerant electrons for iridium oxide Na$_2$IrO$_3$ is derived by an {\it ab initio} downfolding scheme. The model is then reduced to an effective spin model on a honeycomb lattice by the strong coupling expansion. 
Here we show that the {\it ab initio} model contains spin-spin anisotropic exchange terms in addition to the extensively studied Kitaev and Heisenberg exchange interactions, and allows to describe the experimentally observed zigzag magnetic order, interpreted as the state stabilized by the antiferromagnetic coupling of the ferromagnetic chains.  We clarify possible routes to realize quantum spin liquids from existing  Na$_2$IrO$_3$.
\end{abstract}
\pacs{
}
\maketitle
 \paragraph{\tr{Introduction.---}}
Cooperation and competition between strong electron correlations and spin-orbit couplings
have recently attracted much attention.
Iridium oxides offer playgrounds for such an interplay and
indeed exhibit intriguing rich phenomena~\cite{Jackeli,PhysRevLett.105.027204,PhysRevB.83.205101,arXiv:1305.2193}.

Especially, a theoretical prediction~\cite{Jackeli,PhysRevLett.105.027204}
on the possible realization of quantum spin liquid state and Majorana fermion state proven by Kitaev~\cite{AnnalsofPhysics321.2}
as the ground state of an exactly solvable model now called Kitaev model has inspired extensive studies on 
$A_2$IrO$_3$ ($A$= Na or Li ) as a model system to realize the Kitaev spin liquid. 
However, although Na$_2$IrO$_3$ is an insulator (presumably Mott insulator) with the optical gap $\sim 0.35 $ eV~\cite{PhysRevLett.109.266406}, it was shown that Na$_2$IrO$_3$ does not show spin liquid properties experimentally but exhibits a zigzag type magnetic order~\cite{PhysRevLett.108.127204,PhysRevB.85.180403}.

The Kitaev-Heisenberg model on the honeycomb lattice~\cite{Jackeli,PhysRevLett.105.027204,PhysRevLett.108.127203,PhysRevB.84.100406,PhysRevLett.110.097204}
 was further proposed to describe Na$_2$IrO$_3$, which includes isotropic superexchange couplings 
in addition to the Kitaev-type anisotropic nearest-neighbor Ising interactions whose anisotropy axes
depend on the bond directions.
However, it turned out that this model cannot be  
straightforwardly consistent with the zigzag order either.
This discrepancy inspired further studies on suitable low-energy effective
hamiltonians for 
$A_2$IrO$_3$ with $A=$Na or Li.
First, models with further neighbor couplingsi~\cite{PhysRevB.84.180407,PhysRevB.84.024406,PhysRevB.85.180403,PhysRevLett.108.127204}
were studied.
Additional Ising anisotropy~\cite{bhattacharjee2012spin} due to a strong trigonal distortion\textcolor{black}{, which
\mg{actually contradicts the distortions in the experiments~\cite{PhysRevB.85.180403} and in the  {\it ab initio} treatments},}
\mg{was} also examined.
Quasimolecular orbitals~\cite{PhysRevLett.109.197201}, instead of the atomic orbitals 
assumed in the Kitaev-Heisenberg model 
were claimed as a proper choice of the starting point. 
So far the origin of the zigzag type antiferromagnetic order observed for Na$_2$IrO$_3$ and the possible 
route to realize the quantum spin liquid are controversial.

\begin{figure}[ht]
\centering
\includegraphics[width=8.5cm]{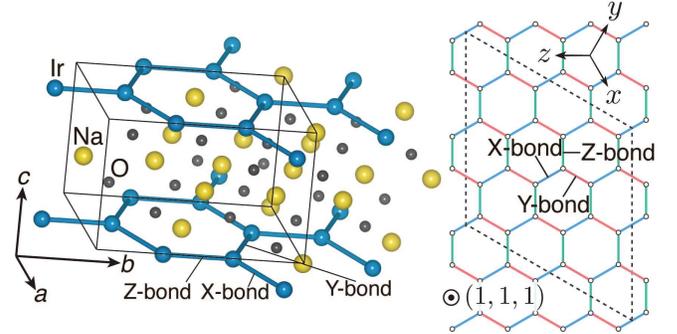}
\caption{(color online):
\mg{Left panel: Crystal structure of Na$_2$IrO$_3$.}
\textcolor{black}{\mg{Right panel:} Honeycomb lattice with $X$-, $Y$-, and $Z$-bonds.
Same colored bonds indicate the same group.
\textcolor{black}{The $x$, $y$, and $z$ \mg{axes in defining} the $t_{2g}$-orbitals
are \mg{illustrated as directions out of the honeycomb plane. The honeycomb plane is then} perpendicular to $(x,y,z)=(1,1,1)$.}
{The dashed boundary represents a 24-site cluster
used later for the exact diagonalization.}
}}
\label{FigH}
\end{figure}

In this Letter, we derive an {\it ab initio} spin model for Na$_2$IrO$_3$ and
show that
trigonal distortions present in Na$_2$IrO$_3$ in addition to the spin-orbit couplings 
hold the key: 
The simplest and realistic
spin model for $A_2$IrO$_3$
\mg{will} turn out to modify the Kitaev-Heisenberg hamiltonian by additional
anisotropic
couplings
as
\textcolor{black}{
\eqsa{
  \hat{H}
  =
  \sum_{\Gamma = X, Y, Z}
  \sum_{\langle \ell,m \rangle \in \Gamma}
  \vec{\hat{S}}_{\ell}^{T}
  \mathcal{J}_{\Gamma}
  \vec{\hat{S}}_{m},
  \label{KHhamiltonian}
}
where $\vec{\hat{S}}_{\ell}^{T}=(\hat{S}^{x}_{\ell},\hat{S}^{y}_{\ell},\hat{S}^{z}_{\ell})$
is a vector of SU(2) spin operators. The exchange couplings are
given in matrices $\mathcal{J}_{\Gamma}$.
The summations are over the nearest-neighbor pairs $\langle \ell,m \rangle$.
The group of bond $\Gamma$ with  $\Gamma$=$X$, $Y$ and $Z$ is defined in Fig.~\ref{FigH}.
The exchange matrices are parametrized as
\eqsa{
&&\mathcal{J}_Z =
\left[
 \begin{array}{ccc}
J   & I_1 & I_2 \\
I_1 & J   & I_2 \\
I_2 & I_2 & K \\
\end{array}
\right],
\mathcal{J}_X =
\left[
\begin{array}{ccc}
K'  & I_{2}'' & I_{2}' \\
I_{2}'' & J''  & I_{1}' \\
I_{2}' & I_{1}' & J' \\
\end{array}
\right],
\nn
&&\mathcal{J}_Y =
\left[
\begin{array}{ccc}
J'' & I_{2}'' & I_{1}' \\
I_{2}'' & K'   & I_{2}' \\
I_{1}' & I_{2}' & J' \\
\end{array}
\right],
}
where we choose a real and symmetric parameterization
\textcolor{black}{by using U(1)- and SU(2)-symmetry of electron wave functions and spin operators, respectively.}
The details of these exchange parameters are described
in the following discussion.}

In addition to the Kitaev coupling \textcolor{black}{$K$ and $K'$,} and XY-type exchange $J$,
magnetic \textcolor{black}{anisotropy} induced by a combination of spin-orbit couplings and trigonal distortions
appears as \textcolor{black}{anisotropic couplings such as \mg{$I_1$ and $I_2$}.
\textcolor{black}{Here we note that \textcolor{black}{our} parameterization of the Kitaev term is
different from \textcolor{black}{that of} Refs.\onlinecite{Jackeli,PhysRevLett.105.027204,PhysRevLett.110.097204}:
The Kitaev term $K$ in the present Letter
\textcolor{black}{corresponds to} $-|K|+J$ in Refs.\onlinecite{Jackeli,PhysRevLett.105.027204,PhysRevLett.110.097204}.}
These anisotropic couplings drastically change candidate quantum phases
and competition among them in Na$_2$IrO$_3$ and related materials.}
With these extensions, we show that the model allows a realistic description of Na$_2$IrO$_3$ and provides a basis for further search of quantum spin liquids.
\textcolor{black}{To achieve quantitative accuracy, we include
the
\textcolor{black}{further} neighbor couplings in our numerical calculations \textcolor{black}{as detailed later}.}

\paragraph{\textcolor{black}{
Ab initio derivation and estimate of  itinerant effective hamiltonian}.---}
To discuss low-energy physics of Na$_2$IrO$_3$, we employ a recently
proposed multi-scale {\it ab initio} scheme for correlated electrons (MACE)~\cite{MiyakeReview2010}: First, we obtain the
global band structure using the density functional theory
(DFT). Second, using a Wannier projection on the Ir
$5d$ $t_{2g}$ target bands, we derive an effective model for the Ir
$5d$ $t_{2g}$ orbitals by the downfolding procedure taking into account the renormalization from the states other than the Ir $5d$ $t_{2g}$ orbitals. 

The global electronic structure was obtained by performing the density functional calculations using the Elk full-potential linearized augmented plane-wave code~\cite{Elk}
\textcolor{black}{with} 
the Perdew-Wang exchange-correlation functional~\cite{PhysRevB.45.13244}. 
The resultant electronic structures agree with the previous DFT results [15] (see Appendix A).

We next constructed the Wannier orbitals from the Ir $t_{2g}$ bands following the same procedure described in Ref.~\cite{PhysRevLett.108.086403}.
\textcolor{black}{One body parameters \textcolor{black}{$t_{\ell,m;a,b}^{\sigma,\sigma'} $} in the low-energy hamiltonian are given by 
 the matrix elements of the Wannier orbitals as
\eqsa{\label{matrix_element}
t_{\ell,m;a,b}^{\sigma,\sigma'}
=
\int
d \vx_1 
w_{\ell a\sigma}^{\ast}(\vx_1)
\hat{H}_{\rm KS}
w_{mb\sigma'}^{\ }(\vx_1)
} with the Kohn-Sham hamiltonian $\hat{H}_{\rm KS}$ and \mg{ the indices for sites $\ell$ and $m$,
orbitals $a$ and $b$, and spins $\sigma$ and $\sigma'$.}}

The effective Coulomb interactions between these orbitals are 
estimated by the constrained random phase approximation (cRPA)~\cite{Aryasetiawan}.
Using the density response code for Elk~\cite{AntoncRPA},
we obtain
the
\textcolor{black}{{\it constrained}} susceptibility
\textcolor{black}{of the noninteracting Kohn-Sham electrons $\chi_0(\vx,\vx',\omega)$
where contribution of particle-hole excitations within the target $t_{2g}$ bands is excluded.}
We then calculate the partially-screened Coulomb interaction
\begin{eqnarray*} \label{eq:w_scr}
  W(\vx,\vx',\omega) =
\frac{1}{|\vx-\vx'|}
+ \int d \vx_1 d \vx_2 \; 
\frac{\chi_{0}(\vx_1,\vx_2,\omega)}{|\vx-\vx_1|} W(\vx_2,\vx',\omega),
\end{eqnarray*}
which yields the \textcolor{black}{Coulomb} interaction
between the Wannier orbitals $w$ as 
\begin{eqnarray*} 
U_{\mathcal{KLMN}} &=& \lim_{\omega \rightarrow 0}\int d \vx_1 d \vx_2 \; w^{*}_{\mathcal{K}}(\vx_1)
w^{*}_{\mathcal{L}}(\vx_2)W(\vx_1,\vx_2,\omega) \times \\
&& w_{\mathcal{M}}(\vx_1)w_{\mathcal{N}}(\vx_2),
\label{eq:u_scr}
\end{eqnarray*} 
where $\mathcal{K},\mathcal{L},\mathcal{M}$, and $\mathcal{N}$ are the combined indices for orbital and site.



\paragraph{\tr{{\it Ab initio} model for $t_{\rm 2g}$ hamiltonian---}}
\textcolor{black}{The} derived multiband 
model consisting of $t_{\rm 2g}$-manifold
of the iridium atoms
is given by the $t_{\rm 2g}$-hamiltonian
\eqsa{
\hat{H}_{t2g}=\hat{H}_{0}+\hat{H}_{\rm tri} + \hat{H}_{\rm SOC} + \hat{H}_{U},
 \label{dHamiltonian}
}
where each decomposed part is determined in the following:
\textcolor{black}{T}he hopping terms are given by 
\textcolor{black}{
\eqsa{ 
  \hat{H}_{0} = 
  \sum_{\ell\neq m}
  \sum_{a,b=xy,yz,zx}
  \sum_{\sigma,\sigma'}
  t_{\ell,m;a,b}^{\sigma\sigma'}
  \left[
  \hat{c}^{\dagger}_{\ell a \sigma}\hat{c}^{\ }_{m b \sigma'} + 
 {\rm h.c.} \right].
}
Here we note that\textcolor{black}{,
among all the hoppings, the dominant terms are the nearest-neighbor hoppings
$t \simeq t_{\ell,m;a,b}^{\sigma\sigma}\simeq t_{\ell,m;b,a}^{\sigma\sigma}$
that satisfy $(a, b)$ = $(zx,xy)$, $(xy,yz)$, or $(yz,zx)$ with $\langle \ell,m\rangle \in X$, $Y$, and $Z$}\textcolor{black}{,
which}
}  
is consistent with the original proposal~\cite{Jackeli} for the Kitaev
couplings.
\textcolor{black}{The $x$-, $y$-, and $z$-axes are illustrated in Fig. 1.}

\textcolor{black}{
The onsite atomic part is derived from Eq.~(\ref{matrix_element}) with $\ell=m$ and can be described as the contribution from
\textcolor{black}{the} trigonal distortion \textcolor{black}{(with orbital-dependent chemical potentials)} and the atomic part of the spin-orbit coupling 
by introducing a vector representation 
$\vec{\hat{c}}^{\dagger}_{\ell}=
(
\hat{c}^{\dagger}_{\ell yz\uparrow},\hat{c}^{\dagger}_{\ell yz\downarrow},
\hat{c}^{\dagger}_{\ell zx\uparrow},\hat{c}^{\dagger}_{\ell zx\downarrow},
\hat{c}^{\dagger}_{\ell xy\uparrow},\hat{c}^{\dagger}_{\ell xy\downarrow}
)$ as
\eqsa{
\hat{H}_{\rm tri}=
\sum_{\ell}
\vec{\hat{c}}^{\dagger}_{\ell}
\left[
\begin{array}{ccc}
-\mu_{yz} & \Delta  & \Delta \\
\Delta & -\mu_{zx} & \Delta \\
\Delta & \Delta & -\mu_{xy} \\
\end{array}
\right]
\hat{\sigma}_0 
\vec{\hat{c}}^{\ }_{\ell},
}
and
\eqsa{
\hat{H}_{\rm SOC}=
\frac{\zeta_{\rm so}}{2}
\sum_{\ell}
\vec{\hat{c}}^{\dagger}_{\ell}
\left[
\begin{array}{ccc}
0 & +i\hat{\sigma}_z & -i\hat{\sigma}_y \\
-i\hat{\sigma}_z & 0 & +i\hat{\sigma}_x \\
+i\hat{\sigma}_y & -i\hat{\sigma}_x & 0 \\
\end{array}
\right]
\vec{\hat{c}}^{\ }_{\ell}.
}
Both the \textcolor{black}{off-diagonal elements of the} spin-independent part $\hat{H}_{\rm tri}$  and the spin-dependent part  $\hat{H}_{\rm SOC}$ can be well described
by a single parameter $\Delta$ and  $\zeta_{\rm so}$, respectively. 
\textcolor{black}{Due to the inherent crystal anisotropy \textcolor{black}{differentiating Ir-Ir bonds along the $b$-axis from others~\cite{arXiv:1312.7437}},
the chemical potential for the $xy$-orbitals, $\mu_{xy}$, is different from
$\mu_{yz}$ and $\mu_{zx}$.}
The symmetry of these terms is slightly broken in the real crystal due to the stacking fault along the $c$-axis and
the locations of other ions. However the deviation is much smaller than \textcolor{black}{0.005} eV.
}

\textcolor{black}{
\textcolor{black}{The Coulomb term expressed by the Wannier orbital basis is well described by
a symmetric form as} 
\eqsa{
&&\hat{H}_{U}
=U\sum_{\ell}\sum_{a=yz,zx,xy}\hat{n}_{\ell a\uparrow}\hat{n}_{\ell a\downarrow}
+\textcolor{black}{\sum_{\ell \neq m}\sum_{a,b}\frac{V_{\ell,m}}{2}\hat{n}_{\ell a}\hat{n}_{m b}}
\nn
&&+
\sum_{\ell }\sum_{a<b}\sum_{\sigma}
\left[
U' \hat{n}_{\ell a\sigma}\hat{n}_{\ell b\overline{\sigma}}
+
(U'-J_{\rm H}) \hat{n}_{\ell a\sigma}\hat{n}_{\ell b\sigma}
\right]
\nn
&&
+J_{\rm H}\sum_{\ell }\sum_{a\neq b}
\left[
\hat{c}^{\dagger}_{\ell a\uparrow}
\hat{c}^{\dagger}_{\ell b\downarrow}
\hat{c}^{\ }_{\ell a\downarrow}
\hat{c}^{\ }_{\ell b\uparrow}
+
\hat{c}^{\dagger}_{\ell a\uparrow}
\hat{c}^{\dagger}_{\ell a\downarrow}
\hat{c}^{\ }_{\ell b\downarrow}
\hat{c}^{\ }_{\ell b\uparrow}
\right],
}
with the local intra-orbital Coulomb repulsion, $U$, 
\textcolor{black}{inter-orbital} Coulomb repulsion, $U'$,
the Hund's rule coupling, $J_{\rm H}$,
the inter-atomic Coulomb repulsion, $V_{\ell,m}$\textcolor{black}{, and
$\hat{n}_{\ell a}=\hat{n}_{\ell a\uparrow}+\hat{n}_{\ell a\downarrow}$}.
The orbital dependences of $U, J_{\rm H}$ and $V$ are 
negligibly small.}

The obtained tight binding parameters are given in Table \ref{TableI}.
We also list the orbital-averaged values of $U, U', J_{\rm H}$ and $V$
obtained by the cRPA.  
We note that $\Delta=-28$ meV \mg{for the $t_{2g}$ model}~\cite{Note1}.
One might think that $\Delta=-28$ meV looks a tiny parameter.
However it is crucial to keep it because it generates relevant anisotropy
\textcolor{black}{illustrated later in Fig. 3.}

\textcolor{black}{
\begin{table}
\textcolor{black}{
\begin{ruledtabular}
\begin{tabular}{c|cccc}
one-body (eV) & $t$ & $\mu_{xy}-\mu_{yz,zx}$ & $\zeta_{\rm so}$ & $\Delta $ \\
\hline
& 0.27 & 0.035 & 0.39 & -0.028 \\ 
\hline
\hline
two-body (eV) & $U$ & $U'$ & $J_{\rm H}$ & $V$ \\
\hline
& 2.72 & 2.09 & 0.23 & 1.1\\
\end{tabular}
\end{ruledtabular}
\caption{One-body and two-body \textcolor{black}{parameters} for $\hat{H}_{t2g}$.
The most relevant hopping parameter $t$, the atomic spin-orbit coupling 
$\zeta$, and the trigonal distortion $\Delta$, are shown for one-body part.
Here, $t$ is for $t_{\ell,m;\xi,\eta}^{\sigma,\sigma}$ for $ \langle \ell,m \rangle $ being the $Z$ bond and its symmetric replacement for $X$ and $Y$ bonds.
As for the two-body parameters,
we list the cRPA results for the local intra-orbital Coulomb repulsion $U$, the Hund's rule coupling $J_{\rm H}$,
and the orbital-independent nearest-neighbor Coulomb repulsion $V$. 
\textcolor{black}{Other \gr{small} one-body parameters are given in Appendix D}.
}
\label{TableI}
}
\end{table}
}



\textcolor{black}{
\begin{table}
\begin{ruledtabular}
\begin{tabular}{l|cccccc}
$\mathcal{J}_Z$ (meV) & $K$ & $J$ & $I_1$ & $I_2$ & & \\ 
\hline
& -30.7 & 4.4 & -0.4 & 1.1 & & \\
\hline
\hline
$\mathcal{J}_{X,Y}$ (meV) & $K'$ & $J'$ & $J''$ & $I'_1$ & $I'_2$ & $I''_2$ \\
\hline
& -23.9 & 2.0 & 3.2 & 1.8 & -8.4 & -3.1 \\
\end{tabular}
\end{ruledtabular}
\caption{
\textcolor{black}{
Nearest-neighbor exchange couplings derived by the strong coupling expansion from the {\it ab initio} $t_{2g}$ model.
}
}
\label{TableIII}
\end{table}
}

\paragraph{\tr{Strong coupling limit, Minimal spin model for $A_2$IrO$_3$.---}}
The {\it ab initio} parameters for the generalized Kitaev-Heisenberg model (\ref{KHhamiltonian}) are derived from $t_{\rm 2g}$ hamiltonian $\hat{H}_{t2g}$ in Eq.(\ref{dHamiltonian}) 
by the second order perturbation theory:
Here we take $\hat{H}_{\rm tri}+\hat{H}_{\rm SOC}+\hat{H}_{U}$ as an unperturbed hamiltonian and
$\hat{H}_0$ as a perturbation.
Since the ground state of $\hat{H}_{\rm tri}+\hat{H}_{\rm SOC}+\hat{H}_{U}$ is degenerate, we employ the standard degenerate perturbation theory.
If we neglect $\Delta$ \textcolor{black}{and $\mu_{a}$ $(a=yz,zx,xy)$} , the lowest Kramers doublets become so-called
$J_{\rm eff}$=1/2 states.
The atomic ground state of an isolated iridium atom is preserved
to be doublet irrespective of the amplitudes of $\Delta$ (see Appendix B), whose
degeneracy is protected by the time-reversal symmetry.
Then the generalized Kitaev-Heisenberg model describing pseudospin degrees of freedom is justified as an effective model
in the ground state as well as at a finite temperature unless it exceeds both of $\Delta$ and $\zeta_{\rm so}$.

\textcolor{black}{The exchange couplings $\mathcal{J}_{Z}$, $\mathcal{J}_{X}$, $\mathcal{J}_{Y}$, 
and \textcolor{black}{further} neighbor couplings}
are derived through the second order 
perturbation theory
\textcolor{black}{by numerically diagonalizing the local part of the hamiltonian $\hat{H}_{\rm tri}+\hat{H}_{\rm SOC}+\hat{H}_{U}$ and
by including all order terms with respect to $\zeta_{\rm so}$ and $\Delta$, irrespective of their amplitudes. \mg{(See Appendix C, D, and E.)}}
\textcolor{black}{
Thus obtained {\it ab initio} values for Na$_2$IrO$_3$ are given in Table \ref{TableIII}.
We remark that \textcolor{black}{$K\sim -30.1$ meV is negative and $J\sim 4.4$ meV is positive for the $Z$-bonds.} 
For numerical calculations, we also include the 2nd and 3rd neighbor couplings
for more accurate {\it ab initio} calculations (see Appendix D).
}

\begin{figure}[htb]
\centering
\includegraphics[width=8.5cm]{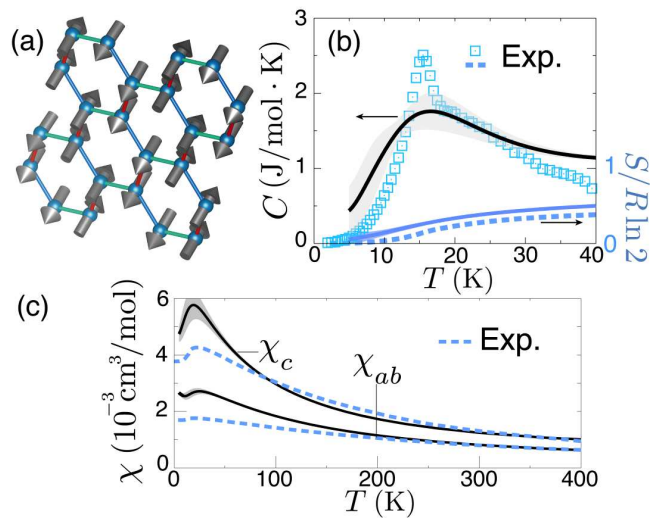}
\caption{(color online):
\textcolor{black}{
Ground state and finite temperature properties of
the generalized Kitaev-Heisenberg model for Na$_2$IrO$_3$
calculated for the 24-site cluster by
using the Lanczos method and thermal pure \textcolor{black}{quantum} states~\cite{PhysRevLett.108.240401}.
(a) Ground state magnetic order determined by applying
tiny local magnetic fields \textcolor{black}{($\sim 10^{-2}$ meV)} at a \textcolor{black}{single} site.
\textcolor{black}{
(b) Temperature-dependence of specific heat $C$ and entropy $S$, which
are consistent with an experiment~\cite{Singh_Gegenwart}.
Shaded area shows uncertainty due to finite size effects~\cite{PhysRevLett.108.240401}.
(c) Temperature-dependence of inplane and out-of-plane magnetic
susceptibilities, which are also
consistent with
the experiment~\cite{Singh_Gegenwart} at high temperatures.
}
}
\label{FigED}
}
\end{figure}

The model (\ref{KHhamiltonian})
with the 
{\it ab initio} 
parameters in Table \ref{TableIII}
\textcolor{black}{together with small and detailed 2nd and 3rd exchange couplings (see Table III in Appendix D)}
was solved by
the exact diagonalization
\textcolor{black}{for a 24-site cluster.}
We also calculate finite temperature properties for the cluster
by using the thermal pure quantum states~\cite{PhysRevLett.108.240401}, which offers
an algorithm similar to the finite-temperature Lanczos~\cite{PhysRevB.49.5065} and earlier works~\cite{Imada_Takahashi}.
\textcolor{black}{They well reproduce the experimentally observed
zigzag magnetic order as the ground state and finite temperature properties.} 
See detailed results in later discussions for \textcolor{black}{Fig.\ref{FigED}}.

Neither large further neighbor exchange couplings~\cite{PhysRevB.84.180407,PhysRevB.84.024406,PhysRevB.85.180403,PhysRevLett.108.127204}
nor antiferromagnetic Kitaev couplings $K>0$\textcolor{black}{~\cite{PhysRevLett.110.097204,arXiv:1310.7940}}
assumed and required to reproduce the experimental zigzag magnetic order in the literature are realistic
in the {\it ab initio} point of view.
In addition, the amplitudes of the anisotropic couplings $I_1$ and $I_2$
comparable with $J$ are crucially important to reproduce the experimental results,
contrary to the assumptions in Refs. \textcolor{black}{\onlinecite{PhysRevLett.110.097204}} and \onlinecite{arXiv:1310.7940}.
\textcolor{black}{The $e_{g}$-orbital degrees of freedom, proposed to change the sign of $K$ in Ref.\onlinecite{PhysRevLett.110.097204}
and neglected in the present Letter,
\textcolor{black}{generate only} minor corrections (see Appendix F).}

\textcolor{black}{
The stabilization of the zigzag order is interpreted as follows:
If we assume the magnetic ordered moment along $(x,y,z)=(1,1,0)$,
the zigzag order is interpreted as
ferromagnetically-ordered chains consisting of the $X$- and $Y$-bonds
(stabilized by $K'$ and $I''_2$),
antiferromagnetically
coupled to each other by the $Z$-bonds with $J$,
which is in contrast to a quantum-chemistry estimate
that neglects $I_2$, $I'_2$, and $I''_2$~\cite{arXiv:1312.7437}.
Indeed, these four exchange couplings, $K\sim K' < 0$, $J>0$, and $I''_2 < 0$, are crucial
to reproduce the zigzag order (see Appendix G).
The alignment
\textcolor{black}{along $(1,1,0)$ assumed here indeed agrees with}
the result of the pinning field analysis (see Appendix H) shown in Fig.\ref{FigED}(a).
\textcolor{black}{
It is also confirmed by
the nearest-neighbor spin-spin correlations,
$
\langle
\hat{S}^{x}_{\ell}
\hat{S}^{x}_{m}
\rangle
=
\langle
\hat{S}^{y}_{\ell}
\hat{S}^{y}_{m}
\rangle
=-0.021,
\langle
\hat{S}^{z}_{\ell}
\hat{S}^{z}_{m}
\rangle
=0.128,
$
for $Z$-bond,
and
$
\langle
\hat{S}^{x}_{\ell}
\hat{S}^{x}_{m}
\rangle
=0.052 (0.098),
\langle
\hat{S}^{y}_{\ell}
\hat{S}^{y}_{m}
\rangle
=0.098 (0.052),
\langle
\hat{S}^{z}_{\ell}
\hat{S}^{z}_{m}
\rangle
=-0.020,
$
for $X$-bond($Y$-bond).
}
}

\paragraph{Comparison with experiments.---}
\textcolor{black}{
\textcolor{black}{Our effective spin model reproduces
not only the zigzag order but
magnetic specific heat and
anisotropic uniform magnetic susceptibilities consistent\textcolor{black}{ly} with experiments,
as shown in Fig. 2(b) and (c).
\textcolor{black}{For the specific heat, our results are consistent without adjustable parameters.}
The uniform magnetic susceptibilities $\chi$ show Curie-Weiss behaviors
and $\chi_{ab} < \chi_{c}$, where $\chi_{ab}$ ($\chi_{c}$) is the inplane (out-of-plane) susceptibility,
which are consistent with experiments.
If we introduce a $g$-factor, $g=1.5$, and \textcolor{black}{anisotropic} van Vleck term, $\chi_0 = 1 \times 10^{-4} {\rm cm^3/mol}$ for $\chi_{c}$,
high-temperature behaviors of $\chi$ are qualitatively reproduced as shown in Fig. 2(c).} 
Here we note that the electron's spin moments are different from those of the effective spin models
depending on the choice of the Kramers doublets, $|\uparrow\rangle$ and $|\downarrow \rangle$ (see Appendix C).
\textcolor{black}{
For the calculation of $\chi$, we project the original Zeeman term to the effective spin basis $\hat{S}_{\ell}^{x,y,z}$ (see Eq.(11) in Appendix C).
It is left for
future studies to relate linear spin wave analysis of our model to 
the inelastic neutron scattering experiment~\cite{PhysRevLett.108.127204}.}
}

\begin{figure}[htb]
\centering
\includegraphics[width=8.75cm]{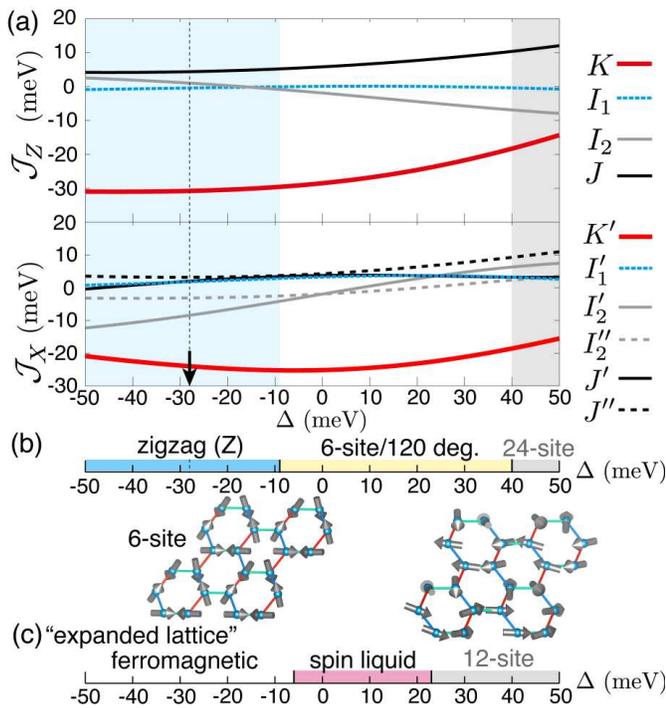}
\caption{(color online):
\textcolor{black}{
(a) $\Delta$-dependence of 
matrix elements of $\mathcal{J}_Z$, $\mathcal{J}_X$
as functions of $\Delta$.
Around the {\it ab initio} values at $\Delta$ ($\sim -28$ meV) listed in Table \ref{TableIII},
$K<0$, $K'<0$, $J>0$, $J'>0$, and $J''>0$ are stably satisfied with gradual dependences on $\Delta$.
(b) Ground state \textcolor{black}{phase} diagram for Na$_2$IrO$_3$ with lattice distortions represented by changes in $\Delta$.
The phase \textcolor{black}{boundaries} are determined by anomalies \textcolor{black}{(peaks signaling \textcolor{black}{continuous} transitions)}
in second derivatives of the exact energy for the 24-site cluster with respect to $\Delta$.
Around the {\it ab initio} parameter $\Delta = -28$ meV, the zigzag order appears.
By increasing $\Delta$, a 6-site unit cell order (or 120$^{\circ}$-structure~\cite{arXiv:1310.7940})
\gr{illustrated in the lower left panel}
and
a 24-site unit cell long-period order (see Appendix H), appear. 
(c) $\Delta$-dependence of the ground state of the generalized Kitaev-Heisenberg model for ``expanded lattices."
Here we neglect the small hopping parameters other than $t$ and take a larger \textcolor{black}{Hund's rule} coupling $J_{\rm H}=0.3$ eV.
Spin liquid phases compete with ferromagnetic states and 12-site unit cell orders
illustrated in the upper right panel (see Appendix H)\textcolor{black}{,
where the phase transitions among them are also interpreted as continuous ones}. 
}
\label{FigJ}
}
\end{figure}

\paragraph{Phase diagram in lattices distorted from Na$_2$IrO$_3$.---}
Now we examine the sensitivity of the ground state for the {\it ab initio} parameter
of Na$_2$IrO$_3$ to perturbations
and search candidates of other quantum states possibly induced
by a thermodynamic control such as pressure or in derivatives of Na$_2$IrO$_3$ such as Na$_{2-x}$Li$_x$IrO$_3$~\cite{PhysRevB.88.220414}. 
\textcolor{black}{Here we choose the trigonal distortion $\Delta$
as an experimentally accessible control parameter.}
First, the $\Delta$-dependence of the exchange couplings is illustrated in Fig.\ref{FigJ}\textcolor{black}{(a)},
where the parameters of the {\it ab initio} $t_{2g}$-hamiltonian other than $\Delta$ are kept unchanged,
\tr{and the exchange couplings are estimated from the same strong coupling expansion by changing $\Delta$.}
The ground state of the generalized Kitaev-Heisenberg model with the $\Delta$-dependent
exchange couplings is shown in Fig.\ref{FigJ}\textcolor{black}{(b)}.

\paragraph{\textcolor{black}{
How to approach spin liquids.}---}
\textcolor{black}{
As already evident in Table.II,
the {\it ab initio} effective spin model for Na$_2$IrO$_3$
is governed by dominant Kitaev-type ferromagnetic exchange couplings.
By expanding the lattice, the spin liquid phase may become accessible:
Expansion of the lattice makes the hopping parameters other than the dominant one $t$
negligible.
In addition, the environment of the iridium atoms approach
the spherical limit
where the intra-orbital Coulomb repulsion $U'$ satisfies $U'=U-2J_{\rm H}$. 
Indeed, when we omit the hopping parameters other than $t$ and increase $J_{\rm H}$ up \textcolor{black}{to} 0.3 eV
to satisfy $U'=U-2J_{\rm H}$, we obtain the spin liquid states
adiabatically connected to the Kitaev's spin liquid as shown in Fig. 3(c).
}

\paragraph{\textcolor{black}{
Summary.---}}
We have shown that the realistic parameter of the {\it ab initio} model for Na$_2$IrO$_3$ reproduces 
the experimentally observed robust zigzag magnetic order, while a quantum spin liquid phase adiabatically connected to the Kitaev spin liquid emerges
when the \textcolor{black}{smaller} trigonal distortion $\Delta$
\textcolor{black}{and expanded lattice constants are satisfied.
In this sense, uniaxial strain to reduce $\Delta$ is helpful as an approach to realize}
the spin liquids.
Clearly further studies are needed: More accurate estimate of the phase diagram of the generalized Kitaev-Heisenberg model is certainly helpful.
More detailed studies by taking account of full quantum fluctuations and the effects of  realistic itinerancy beyond the strong coupling limit are future intriguing issues.
\begin{acknowledgments}
The authors thank Tsuyoshi Okubo for sharing his unpublished data.
Y. N. is supported by the Grant-in-Aid for JSPS Fellows (Grant No.12J08652).
M. K. is supported by Grant-in-Aid for JSPS Fellows (Grant No.12J07338).
R. A. is supported by
Funding Program for World-Leading Innovative R\verb|&|D on Science and
Technology (FIRST program) on ``Quantum Science on Strong Correlation.''
This work is financially
supported by MEXT HPCI Strategic Programs for Innovative Research (SPIRE) (hp130007) and
Computational Materials Science Initiative (CMSI).
Numerical calculation was
partly carried out at the Supercomputer Center,
Institute for Solid State Physics, Univ. of Tokyo.
This work was also supported by Grant-in-Aid for
Scientific Research {(No. 22104010, and No. 223400901)} from MEXT, Japan.
\end{acknowledgments}
\appendix
\section{Appendix A: Details of DFT electronic structure}
\textcolor{black}{
The global electronic structure was obtained by performing the density functional calculations using the Elk full-potential linearized augmented plane-wave code
with the Perdew-Wang exchange-correlation functional. (See Refs.17 and 18.) 
The muffin tin radii ($R_{\rm MT}$) of 1.61, 2.14, and 1.55 bohr for Na, Ir and O were used, respectively. 
The maximum modulus for the reciprocal vectors $K_{\rm max}$ was chosen such that $R^{\rm min}_{\rm MT} K_{\rm max} = 7.0$, 
where $R^{\rm min}_{\rm MT}$ is the smallest $R_{\rm MT}$ in the system. 
}

The present DFT calculation agrees with previously
reported results. Here we show the density of states (DOS) in Fig. 4,
which is consistent with DOS reported in Ref.15 of the main article.
\begin{figure}[ht]
\centering
\includegraphics[width=8cm]{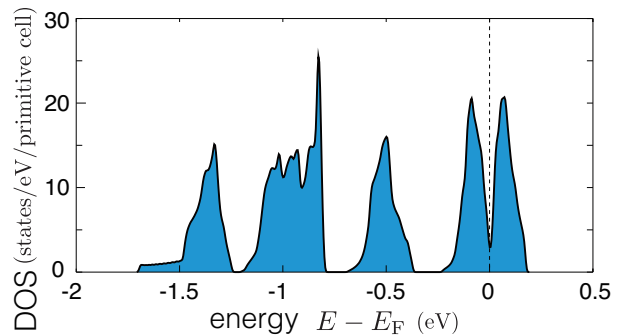}
\caption{
Density of states (DOS) of Na$_2$IrO$_3$ calculated by using
the Elk full-potential linearized augmented plane-wave
code (see the main article and Ref.15).
The crystal structure with the $C2/m$ (No.12) space group
is used, where two iridium atoms exist in the primitive cell.
}
\label{FigDOS}
\end{figure}

\textcolor{black}{
For the calculation of the partially screened Coulomb interaction,
we took 100 unoccupied bands and $4\times 4\times 3$
${\bf k}$ and ${\bf q}$ meshes, and the double Fourier transform of
constrained susceptibility $\chi_0$
was done with
the cutoff of $|{\bf G}+{\bf q}|=5$ (1/a.u.) with ${\bf G}$ being the reciprocal vector.}

\section{Appendix B: Ground state of local hamiltonian}
The eigenstates of the $t_{2g}$-shell of an isolated iridium ion Ir$^{+4}$ is
described by the local hamiltonian $\hat{H}_{\rm tri}+\hat{H}_{\rm SOC}+\hat{H}_{U}$.
For any amplitude of the trigonal distortion $\Delta$,
the atomic ground state with 5 electrons in the $t_{2g}$-shell
is a doublet, as shown in Fig. 5.
The excitation gap among the ground state doublet and the excited doublet
is always larger than $\zeta_{\rm so}=0.39$ eV.
Here, we note that, due to the orbital-dependent chemical potentials
$\mu_{a}$ ($a=xy,yz,zx$), even for $\Delta=0$, there is the finite energy gap
between the first and second excited states.
Therefore, the present pseudo-spin model derived in the main article remains
valid at temperatures \tr{roughly} lower than $\zeta_{\rm so}/k_{\rm B}\simeq \gr{4} \times 10^3$ K.
\begin{figure}[ht]
\centering
\includegraphics[width=7cm]{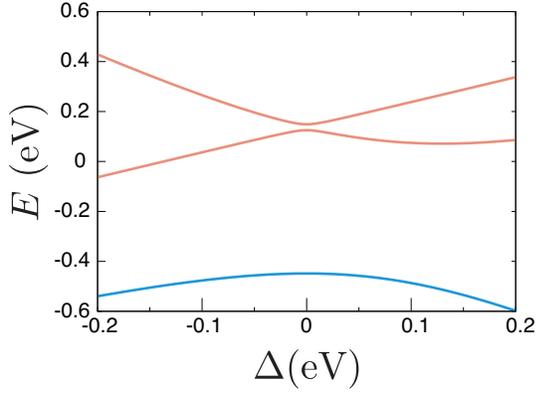}
\caption{
Three doublet eigenstates of the local hamiltonian
$\hat{H}_{\rm tri}+\hat{H}_{\rm SOC}$ with 5 electrons in the $t_{2g}$-shell
as functions of the trigonal distortion $\Delta$.
The Coulomb term $\hat{H}_{U}$ causes a constant shift in these three doublets.
\textcolor{black}{The doublet ground state is shown in the blue curve, while
the red curves represent the excited doublet states.}
}
\label{FigEd5}
\end{figure}

\section{Appendix C: Relationship between physical and effective spins}
\textcolor{black}{Due to the trigonal distortion,
SU(2) rotation and U(1) \textcolor{black}{gauge} transformation of electron wave functions
for the Kramers doublet change the matrix elements of $\mathcal{J}_{\Gamma}$ ($\Gamma=X,Y,Z$),
$\mathcal{J}_2$, and $\mathcal{J}_3$, where, for further neighbor
exchange couplings $\mathcal{J}_2$, and $\mathcal{J}_3$, details are given in the following Appendix D.
In the present Letter,
we choose a Kramers doublet $|\uparrow\rangle$ and $|\downarrow\rangle$ as
\eqsa{
|\uparrow\rangle
&=&
z_1
\hat{c}^{\dagger}_{yz\downarrow}
\hat{c}^{\dagger}_{zx\uparrow}
\hat{c}^{\dagger}_{zx\downarrow}
\hat{c}^{\dagger}_{xy\uparrow}
\hat{c}^{\dagger}_{xy\downarrow}
|0\rangle
\nn
&+&z_2
\hat{c}^{\dagger}_{yz\uparrow}
\hat{c}^{\dagger}_{zx\uparrow}
\hat{c}^{\dagger}_{zx\downarrow}
\hat{c}^{\dagger}_{xy\uparrow}
\hat{c}^{\dagger}_{xy\downarrow}
|0\rangle
\nn
&+&
z_1^{\ast}
\hat{c}^{\dagger}_{yz\uparrow}
\hat{c}^{\dagger}_{yz\downarrow}
\hat{c}^{\dagger}_{zx\downarrow}
\hat{c}^{\dagger}_{xy\uparrow}
\hat{c}^{\dagger}_{xy\downarrow}
|0\rangle
\nn
&-&iz_2
\hat{c}^{\dagger}_{yz\uparrow}
\hat{c}^{\dagger}_{yz\downarrow}
\hat{c}^{\dagger}_{zx\uparrow}
\hat{c}^{\dagger}_{xy\uparrow}
\hat{c}^{\dagger}_{xy\downarrow}
|0\rangle
\nn
&
-&h
\hat{c}^{\dagger}_{yz\uparrow}
\hat{c}^{\dagger}_{yz\downarrow}
\hat{c}^{\dagger}_{zx\uparrow}
\hat{c}^{\dagger}_{zx\downarrow}
\hat{c}^{\dagger}_{xy\downarrow}
|0\rangle
\nn
&+&e^{-i\pi/4}f
\hat{c}^{\dagger}_{yz\uparrow}
\hat{c}^{\dagger}_{yz\downarrow}
\hat{c}^{\dagger}_{zx\uparrow}
\hat{c}^{\dagger}_{zx\downarrow}
\textcolor{black}{\hat{c}^{\dagger}_{xy\uparrow}}
|0\rangle,
}
and
\eqsa{
|\downarrow\rangle
&=&
-z_2^{\ast}
\hat{c}^{\dagger}_{yz\downarrow}
\hat{c}^{\dagger}_{zx\uparrow}
\hat{c}^{\dagger}_{zx\downarrow}
\hat{c}^{\dagger}_{xy\uparrow}
\hat{c}^{\dagger}_{xy\downarrow}
|0\rangle
\nn
&+&z_1^{\ast}
\hat{c}^{\dagger}_{yz\uparrow}
\hat{c}^{\dagger}_{zx\uparrow}
\hat{c}^{\dagger}_{zx\downarrow}
\hat{c}^{\dagger}_{xy\uparrow}
\hat{c}^{\dagger}_{xy\downarrow}
|0\rangle
\nn
&
-&iz_2^{\ast}
\hat{c}^{\dagger}_{yz\uparrow}
\hat{c}^{\dagger}_{yz\downarrow}
\hat{c}^{\dagger}_{zx\downarrow}
\hat{c}^{\dagger}_{xy\uparrow}
\hat{c}^{\dagger}_{xy\downarrow}
|0\rangle
\nn
&+&z_1
\hat{c}^{\dagger}_{yz\uparrow}
\hat{c}^{\dagger}_{yz\downarrow}
\hat{c}^{\dagger}_{zx\uparrow}
\hat{c}^{\dagger}_{xy\uparrow}
\hat{c}^{\dagger}_{xy\downarrow}
|0\rangle
\nn
&
-&e^{+i\pi/4}f
\hat{c}^{\dagger}_{yz\uparrow}
\hat{c}^{\dagger}_{yz\downarrow}
\hat{c}^{\dagger}_{zx\uparrow}
\hat{c}^{\dagger}_{zx\downarrow}
\hat{c}^{\dagger}_{xy\downarrow}
|0\rangle
\nn
&-&h
\hat{c}^{\dagger}_{yz\uparrow}
\hat{c}^{\dagger}_{yz\downarrow}
\hat{c}^{\dagger}_{zx\uparrow}
\hat{c}^{\dagger}_{zx\downarrow}
\textcolor{black}{\hat{c}^{\dagger}_{xy\uparrow}}
|0\rangle,
}
where $z_1$, $z_2$, $f$ and $h$ are coefficients of the linear combinations.
In the above parameterization of the Kramers doublets,
the $J_{\rm eff}=1/2$-state is represented by taking the coefficients
$z_1 = (1-i)/\sqrt{6}$, $z_2=h=0$, and $f=1/\sqrt{6}$.
Our choice for the above Kramers doublet
give us real number elements in the nearest-neighbor exchange coupling
along the $Z$-bond, $\mathcal{J}_Z$.
}

\textcolor{black}{For the {\it ab initio} model, we choose
the parameter set $(z_1,z_2,f,h)$ \gr{that diagonalizes}
the $z$-component of the reduced
spin operators for the $t_{\rm 2g}$-manifold defined as
\eqsa{
\left[\widetilde{S}^{\alpha}_{\rm tot}\right]_{\sigma\sigma'}=
\langle \sigma|
\sum_{a}\sum_{\sigma,\sigma'} \hatd{c}{a \sigma} \hat{\sigma}^{\alpha}\hatn{c}{a \sigma'}/2
|\sigma'\rangle,
}
where $\alpha=x,y,z$.
By using the resultant Kramers doublet,
the total magnetic moment consisting of the reduced spin and angular momentum operators, $\widetilde{S}^{\alpha}_{\rm tot}$ and
$\widetilde{L}^{\alpha}_{\rm tot}$, for the $t_{\rm 2g}$-manifold
is expressed by the SU(2) operators $\hat{S}^{\alpha}$.
For the {\it ab initio} model, the total magnetic moment
is given as
\eqsa{
\left[
\begin{array}{c}
2\widetilde{S}^x_{\rm tot}-\widetilde{L}^x_{\rm tot}\\
2\widetilde{S}^y_{\rm tot}-\widetilde{L}^y_{\rm tot}\\
2\widetilde{S}^z_{\rm tot}-\widetilde{L}^z_{\rm tot}\\
\end{array}
\right]
=
2
\left[
\begin{array}{ccc}
-0.07 & +0.94 & -0.24 \\ 
+0.94 & -0.07 & +0.24 \\ 
-0.07 & +0.07 & +1.07 \\ 
\end{array}
\right]
\left[
\begin{array}{c}
\hat{S}^x\\
\hat{S}^y\\
\hat{S}^z\\
\end{array}
\right].\nonumber\\
}
}
\textcolor{black}{For the calculation of the uniform magnetic susceptibilities $\chi$,
we reduce the original Zeeman term, 
\eqsa{
-\mu_{\rm B}(2\vec{\widetilde{S}}-\vec{\widetilde{L}})\cdot \vec{B},\nonumber
}
to the effective spin basis $\hat{S}_{\ell}^{x,y,z}$ through Eq.(12).}

\section{Appendix D:
Further neighbor exchange couplings and details of the hopping parameters}
For quantitative accuracy, we include dominant 2nd and 3rd neighbor exchange couplings
represented by $\hat{H}'$
for our numerical calculations:
\eqsa{
  \hat{H}'=
  \sum_{\langle \ell,m \rangle' \in Z_{2nd}}
  \vec{\hat{S}}_{\ell}^{T}
  \mathcal{J}_2
  \vec{\hat{S}}_{m}
  +
  \sum_{\langle \ell,m \rangle''}
  \vec{\hat{S}}_{\ell}^{T}
  \mathcal{J}_3
  \vec{\hat{S}}_{m},
}
\gr{where} further exchange couplings are
given in matrices $\mathcal{J}_2$, and $\mathcal{J}_3$.
The summations are over
the second neighbor pairs $\langle \ell,m \rangle'$, and the third neighbor pairs $\langle \ell,m \rangle''$.
For the 2nd neighbor pairs, exchange couplings are finite if they belong to the group of 2nd
neighbor bonds perpendicular to the $Z$-bond, $Z_{\rm 2nd}$.
These exchange matrices are parametrized as
\eqsa{
\mathcal{J}_2 =
\left[
\begin{array}{ccc}
J^{\rm (2nd)}   & I_1^{\rm (2nd)} & I_2^{\rm (2nd)} \\
I_1^{\rm (2nd)} & J^{\rm (2nd)}   & I_2^{\rm (2nd)} \\
I_2^{\rm (2nd)} & I_2^{\rm (2nd)} & K^{\rm (2nd)} \\
\end{array}
\right],
}
and
\eqsa{
\mathcal{J}_3 =
\left[
\begin{array}{ccc}
J^{\rm (3rd)} & 0 & 0 \\
0 & J^{\rm (3rd)} & 0 \\
0 & 0 & J^{\rm (3rd)} \\
\end{array}
\right],
}
\gr{where the obtained parameters are given in Table III.
The \textcolor{black}{all of the matrix elements of the} 2nd and 3rd neighbor exchange couplings for other bonds,
Dzyaloshinskii-Moriya-type couplings,
as well as the couplings for even further neighbor bonds are
\textcolor{black}{smaller than 1 meV and neglected}.}

For derivation of these exchange parameters,
we use detailed {\it ab initio} hopping parameters
summarized in Table IV.
\textcolor{black}{
\begin{table}
\begin{ruledtabular}
\begin{tabular}{l|cccc}
$\mathcal{J}_2$ (meV) & $K^{\rm (2nd)}$ & $J^{\rm (2nd)}$ & $I_1^{\rm (2nd)}$ & $I_2^{\rm (2nd)}$ \\ 
\hline
& -1.2 & -0.8 & 1.0 & -1.4 \\
\hline
\hline
$\mathcal{J}_3$ (meV) & $J^{\rm (3rd)}$ & & & \\
\hline
& 1.7 & & & \\
\end{tabular}
\end{ruledtabular}
\caption{
\textcolor{black}{
Second and third neighbor
exchange couplings derived by the strong coupling expansion from the {\it ab initio} $t_{2g}$ model.
}
}
\label{TableIV}
\end{table}
}

\begin{table}[htb]
\begin{ruledtabular}
\begin{tabular}{l|rrrrrr}
$Z$& $yz\uparrow$ & $yz\downarrow$ & $zx\uparrow$ & $zx\downarrow$ & $xy\uparrow$ & $xy\downarrow$ \\
\hline
$yz\uparrow$ &  31+0i &   0+0i & 273+8i &   4+4i & -16-2i &  10+39i \\
$yz\downarrow$ &   0+0i &  31+0i &  -4+4i & 273-8i & -10+39i & -16+2i \\
$zx\uparrow$ & 273-8i &  -4-4i &  31+0i &   0+0i & -16+2i & -39-10i \\
$zx\downarrow$ &   4-4i & 273+8i &   0+0i &  31+0i &  39-10i & -16-2i \\
$xy\uparrow$ & -16+2i & -10-39i & -16-2i &  39+10i &  43+0i &   0+0i \\
$xy\downarrow$ &  10-39i & -16-2i & -39+10i & -16+2i &   0+0i &  43+0i \\
\hline
\hline
$X$& $yz\uparrow$ & $yz\downarrow$ & $zx\uparrow$ & $zx\downarrow$ & $xy\uparrow$ & $xy\downarrow$ \\
\hline
$yz\uparrow$ &  -7+0i &   0+0i & -18-11i & -36+3i & -25+39i &  11-2i \\
$yz\downarrow$ &   0+0i &  -7+0i &  36+3i & -18+11i & -11-2i & -25-39i \\
$zx\uparrow$ & -18+11i &  36-3i &  36+0i &   0+0i & 276+5i &   5+8i \\
$zx\downarrow$ & -36-3i & -18-11i &   0+0i &  36+0i &  -5+8i & 276-5i \\
$xy\uparrow$ & -25-39i & -11+2i & 276-5i &  -5-8i &  38+0i &   0+0i \\
$xy\downarrow$ &  11+2i & -25+39i &   5-8i & 276+5i &   0+0i &  38+0i \\
\hline
\hline
$Z_{\rm 2nd}$& $yz\uparrow$ & $yz\downarrow$ & $zx\uparrow$ & $zx\downarrow$ & $xy\uparrow$ & $xy\downarrow$ \\
\hline
$yz\uparrow$ &   0-3i &   0+4i & -85+0i &  -2-2i &  12-2i &  -1+9i \\
$yz\downarrow$ &   0+4i &   0+3i &   2-2i & -85+0i &   1+9i &  12+2i \\
$zx\uparrow$ & -30+1i &  -2-2i &   0-3i &   4+0i & -20+3i & -10+0i \\
$zx\downarrow$ &   2-2i & -30-1i &  -4+0i &   0+3i &  10+0i & -20-3i \\
$xy\uparrow$ & -20+3i &   0-10i &  12-2i &   9 -1i &  -2+6i &   0+0i \\
$xy\downarrow$ &   0-10i & -20-3i &  -9-1i &  12+2i &   0+0i &  -2-6i \\
\hline
\hline
$Z_{\rm 3rd}$& $yz\uparrow$ & $yz\downarrow$ & $zx\uparrow$ & $zx\downarrow$ & $xy\uparrow$ & $xy\downarrow$ \\
\hline
$yz\uparrow$ & -10+0i &   0+0i & -13+0i &  -1-1i &  18+0i &   0+4i \\
$yz\downarrow$ &   0+0i & -10+0i &   1-1i & -13+0i &   0+4i &  18+0i \\
$zx\uparrow$ & -13+0i &   1+1i & -10+0i &   0+0i &  18+0i &  -4+0i \\
$zx\downarrow$ &  -1+1i & -13+0i &   0+0i & -10+0i &   4+0i &  18+0i \\
$xy\uparrow$ &  18+0i &   0-4i &  18+0i &   4+0i & -37+0i &   0+0i \\
$xy\downarrow$ &   0-4i &  18+0i &  -4+0i &  18+0i &   0+0i & -37+0i \\
\end{tabular}
\end{ruledtabular}
\caption{
Detailed hopping parameters for nearest-, 2nd, and 3rd neighbor pairs of the iridium sites.
The unit of the hopping parameters is given by meV.
For the nearest-neighbor hoppings, the hoppings along the $Z$-bond and $X$-bond are shown,
where the hoppings along the $Y$-bond is obtained by exchanging indices in the hoppings
along the $X$-bond. 
The directions of these hopping processes are illustrated in Fig. 8(a) later. 
}
\label{TableV}
\end{table}

\section{Appendix E: Second order perturbation}
In the present Letter, we derive 
a generalized Kitaev-Heisenberg model
by employing second-order degenerated perturbation theory
from a strong coupling limit:
We perform the perturbation by taking $\hat{H}_{\rm tri}+\hat{H}_{\rm SOC}+\hat{H}_{U}$
as the unperturbed hamiltonian and $\hat{H}_0$ as the perturbation.

As clarified in the Appendix D,
the ground state of the local part of the unperturbed hamiltonian
$\hat{H}_{\rm tri}+\hat{H}_{\rm SOC}+\hat{H}_{U}$ 
is a Kramers doublet, \tr{but, strictly speaking, it deviates} from the so-called
$J_{\rm eff}=1/2$ state if $\Delta\neq 0$.
We assign pseudo-spin degree of freedom
to this doublet.

For illustrative purpose,
we focus on a set of nearest-neighbor sites, the $\ell$-th and $m$-th sites.
Then we calculate the perturbation energy through the second order processes as
\eqsa{
E^{(2)}_{\sigma_1,\sigma_2 ; \sigma_3, \sigma_4}=
\bra{m \sigma_2 }
\bra{\ell \sigma_1 }
\hat{H}_0
\sum_{n}
\frac{|n\rangle\rangle\langle\langle n|}{E_n - E_0 }
\hat{H}_0
\ket{\ell \sigma_3 }
\ket{m \sigma_4 },\nonumber\\
} 
where $\sigma_j=\uparrow,\downarrow$ $(j=1,2,3,4)$ is a pseudo-spin index, and
$|n\rangle\rangle$ is an intermediate eigenstate of $\hat{H}_{\rm tri}+\hat{H}_{\rm SOC}+\hat{H}_{U}$
with 4 and 6 electrons at the $\ell$-th and $m$-th site, respectively, or
6 and 4 electrons at the $\ell$-th and $m$-th site, respectively.
Here $E_0$ is the ground-state energy of the two sites with 5 electrons per site
and $E_n$ is an energy eigenvalues of an intermediate state of the two sites.  
The eigenstates $|n\rangle\rangle$ and eigenvalues $E_n$ are obtained by
numerically diagonalizing $\hat{H}_{\rm tri}+\hat{H}_{\rm SOC}+\hat{H}_{U}$.

From the perturbation energy $E^{(2)}_{\sigma_1,\sigma_2 ; \sigma_3, \sigma_4}$,
we obtain the exchange couplings as follows.
If we assume the bond connecting the $\ell$-th and $m$-th sites is a $Z$-bond,
the exchange couplings are given for the minimal spin model for $A_2$IrO$_3$ as
\eqsa{
  K&=&+2\left[
     E^{(2)}_{\sigma,\sigma ; \sigma, \sigma}-E^{(2)}_{\sigma,\overline{\sigma} ; \sigma, \overline{\sigma}}
     \right],\\
  J&=&+2E^{(2)}_{\sigma,\overline{\sigma} ; \overline{\sigma}, \sigma},\\
  I_1&=&-2{\rm Im}\left\{E^{(2)}_{\uparrow,\uparrow;\downarrow,\downarrow}\right\}=
        +2{\rm Im}\left\{E^{(2)}_{\downarrow,\downarrow;\uparrow,\uparrow}\right\},\\
  I_2&=&+4{\rm Re}\left\{E^{(2)}_{\uparrow,\uparrow;\uparrow,\downarrow} \right\}
      = -4{\rm Im}\left\{E^{(2)}_{\uparrow,\uparrow;\uparrow,\downarrow} \right\}\nn
     &=&+4{\rm Re}\left\{E^{(2)}_{\uparrow,\uparrow;\downarrow,\uparrow} \right\}
      = -4{\rm Im}\left\{E^{(2)}_{\uparrow,\uparrow;\downarrow,\uparrow} \right\}\nn
     &=&-4{\rm Re}\left\{E^{(2)}_{\downarrow,\downarrow;\uparrow,\downarrow} \right\}
      = -4{\rm Im}\left\{E^{(2)}_{\downarrow,\downarrow;\uparrow,\downarrow} \right\}\nn
     &=&-4{\rm Re}\left\{E^{(2)}_{\downarrow,\downarrow;\downarrow,\uparrow} \right\}
      = -4{\rm Im}\left\{E^{(2)}_{\downarrow,\downarrow;\downarrow,\uparrow} \right\}.
}

For the 2nd and 3rd neighbor bond, the matrix elements of the exchange couplings $\mathcal{J}_2$ and $\mathcal{J}_3$ 
as functions of the trigonal distortion $\Delta$ are calculated by the 2nd order perturbation as shown in Fig. 6.
\begin{figure}[ht]
\centering
\includegraphics[width=8.5cm]{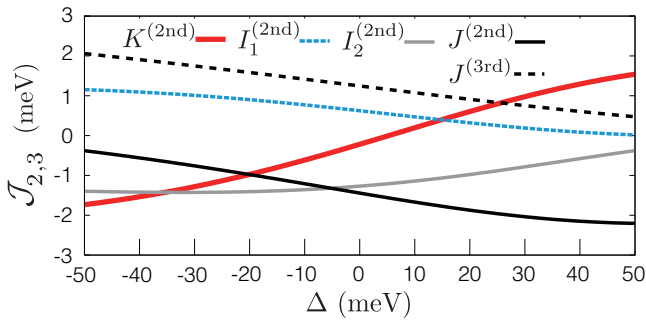}
\caption{
Exchange couplings for 2nd and 3rd neighbor bond as functions of $\Delta$.
}
\label{FigJ23}
\end{figure}

\section{Appendix F: An estimate of contributions from $e_g$-orbitals}
\textcolor{black}{
\textcolor{black}{In Ref.11, it was proposed that the}
$t_{2g}$-$e_{g}$ hoppings
\textcolor{black}{might play an important role in determining the signs of the Kitaev couplings.
However, we show that they give}
only small corrections to the Kitaev couplings, $K$ and $K'$ by employing
the estimation of Ref.11 combined with parameters expected from our $ab$ $initio$ parameters for the $t_{2g}$-manifold.
Following Ref.11, we employ the formula, $(4/9)(\widetilde{t}/\widetilde{U})^2 \widetilde{J}_{\rm H}$ for the contribution of
the $t_{2g}$-$e_{g}$ hoppings to the Kitaev couplings.
To evaluate the formula, we choose the $t_{2g}$-$e_{g}$ hoppings $\widetilde{t}=2t=$0.54 eV,
the Hund's \textcolor{black}{rule} coupling among the $t_{2g}$- and $e_{g}$-orbitals, $\widetilde{J}_{\rm H}=J_{\rm H}=$0.23 eV, and the excitation energy
of intermediate states for the 2nd order perturbation process, $\widetilde{U}$=$U-V+\Delta_{e_{g}t_{2g}}$=2.7-1.1+3=4.6 eV,
where the crystal field splitting between the $e_{g}$ and $t_{2g}$, $\Delta_{e_{g}t_{2g}}$=3 eV, is estimated from the $ab$ $initio$
band structure given in Ref.15.
Then, the correction to the Kitaev coupling from the $t_{2g}$-$e_{g}$ hoppings is estimated as $(4/9)(\widetilde{t}/\widetilde{U})^2 \widetilde{J}_{\rm H}$
=1.4 meV, which is less than 6 percent of our $ab$ $initio$ estimate of $K$ and $K'$.
Therefore, the $t_{2g}$-$e_{g}$ hoppings introduce minor quantitative corrections and do not change the sign of the Kitaev couplings,
which justifies our $t_{2g}$ hamiltonian as a proper effective low-energy hamiltonian.}

\section{Appendix G: Stabilization of zigzag magnetic orders}
As explained in the main article,
the zigzag order found as the ground state of the {\it ab initio} model
is interpreted as antiferromagnetically coupled ferromagnetic chains,
which is stabilized by the three exchange couplings,
$K\sim K'<0$, $J>0$, and $I''_2 <0 $. 
\textcolor{black}{As explained later, another exchange coupling $I'_2$, which seemingly dominates $I''_2$,
is irrelevant to the zigzag (Z) order.}
To demonstrate this, we show that the zigzag order which is adiabatically
connected to that of the {\it ab initio} model is indeed realized in a simplified model with
these three exchange couplings,
$K\sim K'<0$, $J>0$, and $I''_2 <0 $, defined as
\begin{figure}[tb]
\centering
\includegraphics[width=7.5cm]{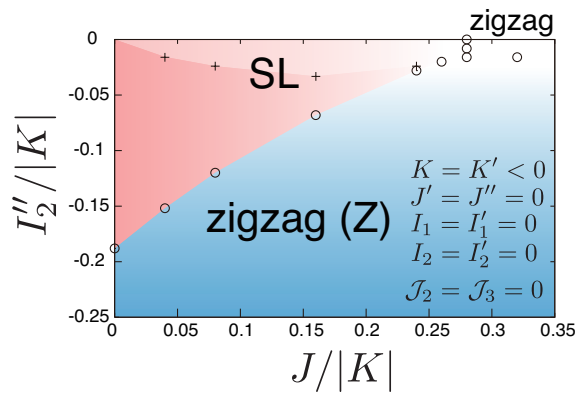}
\caption{
Phase diagram for the ground states of the simplified model, $\hat{H}_{KJI''_2}$,
with $K=K'<0$, $J>0$, and $I''_2<0$, defined in Eqs. (\ref{Hsimp}) and (\ref{Jsimp}).
There are manly two kind of phases: spin liquid (SL) phases adiabatically connected to
the Kitaev limit ($J=I''_2=0$) (red region) and zigzag order phases (blue region).
Especially, the phase denoted as zigzag (Z) shows the Bragg peak located at the
momentum $(0,1)$ shown as blue diamonds in Fig. 8(b).
For small amplitude of $I''_2$ and $J/|K|>0.28$,
zigzag-type magnetic \tr{correlations are enhanced but are accompanied} with additional
peaks \tr{as well} corresponding to other alignments of zigzag orders,
\tr{which makes a clear determination of the phase difficult}.
For determination of \tr{other} phase boundaries, the second derivatives of
the ground state energy with respect to $J$ or $I''_2$ are employed.
To identify the phase without ambiguity,
we also check the calculated results on Bragg peaks in the magnetic structure factors.
Open circles indicate continuous phase transitions (or peaks in the second derivatives
of the ground state energy), and crosses indicate first order phase transitions
(or cusps \tr{(level crossings)} in the ground state energy). 
}
\label{Figphase}
\end{figure}
\eqsa{
  \hat{H}_{KJI''_2}
  =
  \sum_{\Gamma = X, Y, Z}
  \sum_{\langle \ell,m \rangle \in \Gamma}
  \vec{\hat{S}}_{\ell}^{T}
  \mathcal{J}_{\Gamma}^{(0)}
  \vec{\hat{S}}_{m},\label{Hsimp}
}
where the matrices of the exchange couplings are given by
\eqsa{
&&\mathcal{J}_Z^{(0)} =
\left[
 \begin{array}{ccc}
J   & 0 & 0 \\
0 & J   & 0 \\
0 & 0 & K \\
\end{array}
\right],
\mathcal{J}_X^{(0)} =
\left[
\begin{array}{ccc}
K  & I_{2}'' & 0 \\
I_{2}'' & 0  & 0 \\
0 & 0 & 0 \\
\end{array}
\right],
\nn
&&\mathcal{J}_Y^{(0)} =
\left[
\begin{array}{ccc}
0 & I_{2}'' & 0 \\
I_{2}'' & K   & 0 \\
0 & 0 & 0 \\
\end{array}
\right].\label{Jsimp}
}

The ground states of the simplified model $\hat{H}_{KJI''_2}$
are summarized as the phase diagram in Fig. 7. 
The three exchange couplings, $K\sim K' <0$, $J>0$, and $I''_2<0$,
indeed stabilize the zigzag order,
and the positions of the Bragg peaks are identical with those of
the zigzag-ordered phase obtained by the {\it ab initio} model,
which is denoted as zigzag (Z) in Fig. 8.
\tr{We have confirmed that the
present zigzag ordered phase is adiabatically connected with the zigzag order found in the {\it ab initio} model.} 

\textcolor{black}{Here we note that the present model with the three exchange couplings, $K$, $I''_2$, and $J$,
is given by straightforward simplification of the $ab$ $initio$ model: First, we discard small exchange couplings,
$I_1$, $I_2$, $J$, $I'_1$, $\mathcal{J}_2$, and $\mathcal{J}_3$, with amplitudes less than 3 meV.
Then, we drop the sub-dominant Heisenberg term $J''$. As explained below, $I'_2$ is irrelevant for the zigzag (Z)
order and, therefore, is dropped in the present three-exchange-coupling model.}

\textcolor{black}{By averaging the $ab$ $initio$ Kitaev couplings as $(K+2K')/3 \sim$ 26 meV, the parameter set
$I''_2/|K|=-0.12$ and $J/|K|=0.17$ corresponds to the $ab$ $initio$ model and gives the zigzag (Z) order as shown in Fig. 7.
For the parameter set,
when we introduce additional $I'_2$ up to $I'_2/|K|=-0.32$ that corresponds to the $ab$ $initio$ value,
we confirm that the ground state remains the zigzag (Z) order for the 24-site cluster.}

\textcolor{black}{We also note that, in the simplified three-exchange-coupling model,
the difference between $\mathcal{J}_{Z}^{(0)}$ and $\mathcal{J}_{X}^{(0)}$
(or $\mathcal{J}_{Y}^{(0)}$) inherits from
the anisotropic crystal structure
that differentiates Ir-Ir bonds along the $b$-axis
from other bonds.
If we take an isotropic exchange couplings for $\mathcal{J}_{Z}^{(0)}$, $\mathcal{J}_{X}^{(0)}$,
and $\mathcal{J}_{Y}^{(0)}$, we do not obtain the zigzag (Z) states as the ground states\textcolor{black}{,
which indicates that the inherent anisotropy plays an important role in
realizing the zigzag (Z) order}.}

\section{Appedix H: Magnetic Bragg peaks and pinning field analysis}

\begin{table}[htb]
\begin{ruledtabular}
\begin{tabular}{llll}
zigzag (Z) & 6-site/120$^{\circ}$ & 24-site & 12-site\\
\hline
$(0,1)$ & $(1/3, 1)$ & $(1/6,1/2)$ & $(1/3,0)$\\
\end{tabular}
\end{ruledtabular}
\caption{
\textcolor{black}{
List of \gr{the momenta at which} the dominant peaks \gr{appear} in the magnetic structure factors
calculated for each magnetic ordered phases
in the phase diagrams, Fig. 3(b) and (c) of the main article.
Momenta are defined in a two dimensional Brillouin zone
that is used in Ref.7 of the main article and Fig. 8.
}
}
\label{TableIV}
\end{table}

\textcolor{black}{Peaks in spin structure factors, which may correspond to the magnetic Bragg peaks
in the thermodynamic limit,}
are primarily used
to determine magnetic orders in the exact diagonalization.
\textcolor{black}{The 2D unit cell in the honeycomb layer of Na$_2$IrO$_3$ and the peaks of the spin structure factors are shown in Fig. 8.}
In Table V, the magnetic Bragg peak \gr{positions in the two-dimensional Brillouin zone} are summarized.

However, anisotropy in alignments of ordered moments
cannot be determined by the magnetic Bragg spots.
Therefore we applied tiny local magnetic fields ($\sim 10^{-2}$ meV)
to break symmetries and pin down the magnetic order pattern
during the Lanczos steps.
In Fig. 2 and Fig. 3 of the main article,
normalized 
induced moments 
are illustrated.
\begin{figure}[htb]
\centering
\includegraphics[width=6.5cm]{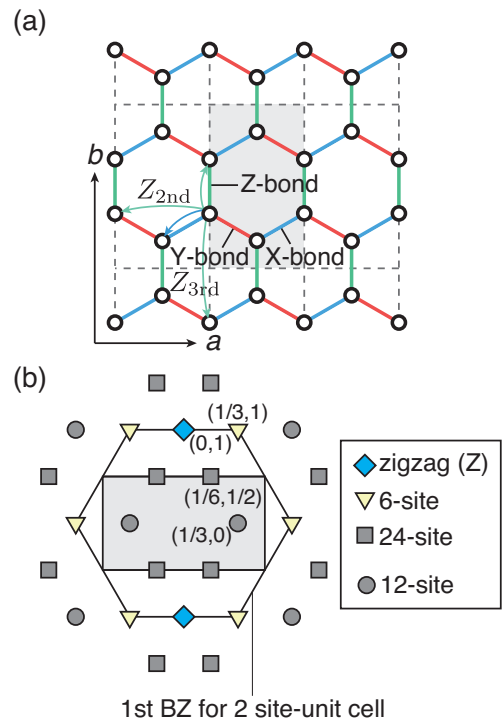}
\caption{
\textcolor{black}{
(a) Experimental unit cell for honeycomb planes of iridium atoms.
The horizontal and vertical axes represent the $a$- and $b$-axis
of Na$_2$IrO$_3$.
The shaded rectangular area illustrates an unit cell
consisting of 4 iridium atoms.
The arrows illustrate the direction of the hoppings summarized in Table IV. 
(b) Peaks in spin structure factors calculated by
using the Lanczos method for the 24-site cluster.
The shaded rectangular area illustrates the 1st Brillouin zone (BZ)
of Na$_2$IrO$_3$.
The hexagon represents the 1st Brillouin zone of the ideal honeycomb lattice
whose unit cell contains two sites. 
}
}
\label{FigBragg}
\end{figure}
\bibliographystyle{apsrev}
\bibliography{Ir213}

\begin{thebibliography}{30}
\expandafter\ifx\csname natexlab\endcsname\relax\def\natexlab#1{#1}\fi
\expandafter\ifx\csname bibnamefont\endcsname\relax
  \def\bibnamefont#1{#1}\fi
\expandafter\ifx\csname bibfnamefont\endcsname\relax
  \def\bibfnamefont#1{#1}\fi
\expandafter\ifx\csname citenamefont\endcsname\relax
  \def\citenamefont#1{#1}\fi
\expandafter\ifx\csname url\endcsname\relax
  \def\url#1{\texttt{#1}}\fi
\expandafter\ifx\csname urlprefix\endcsname\relax\def\urlprefix{URL }\fi
\providecommand{\bibinfo}[2]{#2}
\providecommand{\eprint}[2][]{\url{#2}}

\bibitem[{\citenamefont{Jackeli and Khaliullin}(2009)}]{Jackeli}
\bibinfo{author}{\bibfnamefont{G.}~\bibnamefont{Jackeli}} \bibnamefont{and}
  \bibinfo{author}{\bibfnamefont{G.}~\bibnamefont{Khaliullin}},
  \bibinfo{journal}{Phys. Rev. Lett.} \textbf{\bibinfo{volume}{102}},
  \bibinfo{pages}{017205} (\bibinfo{year}{2009}).

\bibitem[{\citenamefont{Chaloupka et~al.}(2010)\citenamefont{Chaloupka,
  Jackeli, and Khaliullin}}]{PhysRevLett.105.027204}
\bibinfo{author}{\bibfnamefont{J.}~\bibnamefont{Chaloupka}},
  \bibinfo{author}{\bibfnamefont{G.}~\bibnamefont{Jackeli}}, \bibnamefont{and}
  \bibinfo{author}{\bibfnamefont{G.}~\bibnamefont{Khaliullin}},
  \bibinfo{journal}{Phys. Rev. Lett.} \textbf{\bibinfo{volume}{105}},
  \bibinfo{pages}{027204} (\bibinfo{year}{2010}).

\bibitem[{\citenamefont{Wan et~al.}(2011)\citenamefont{Wan, Turner, Vishwanath,
  and Savrasov}}]{PhysRevB.83.205101}
\bibinfo{author}{\bibfnamefont{X.}~\bibnamefont{Wan}},
  \bibinfo{author}{\bibfnamefont{A.~M.} \bibnamefont{Turner}},
  \bibinfo{author}{\bibfnamefont{A.}~\bibnamefont{Vishwanath}},
  \bibnamefont{and} \bibinfo{author}{\bibfnamefont{S.~Y.}
  \bibnamefont{Savrasov}}, \bibinfo{journal}{Phys. Rev. B}
  \textbf{\bibinfo{volume}{83}}, \bibinfo{pages}{205101}
  (\bibinfo{year}{2011}).

\bibitem[{\citenamefont{Witczak-Krempa et~al.}()\citenamefont{Witczak-Krempa,
  Gang, Kim, and Balents}}]{arXiv:1305.2193}
\bibinfo{author}{\bibfnamefont{W.}~\bibnamefont{Witczak-Krempa}},
  \bibinfo{author}{\bibfnamefont{C.}~\bibnamefont{Gang}},
  \bibinfo{author}{\bibfnamefont{Y.-B.} \bibnamefont{Kim}}, \bibnamefont{and}
  \bibinfo{author}{\bibfnamefont{L.}~\bibnamefont{Balents}},
  \eprint{arXiv:1305.2193}.

\bibitem[{\citenamefont{Kitaev}(2006)}]{AnnalsofPhysics321.2}
\bibinfo{author}{\bibfnamefont{A.}~\bibnamefont{Kitaev}},
  \bibinfo{journal}{Annals Phys.} \textbf{\bibinfo{volume}{321}},
  \bibinfo{pages}{2} (\bibinfo{year}{2006}).

\bibitem[{\citenamefont{Comin et~al.}(2012)\citenamefont{Comin, Levy, Ludbrook,
  Zhu, Veenstra, Rosen, Singh, Gegenwart, Stricker, Hancock
  et~al.}}]{PhysRevLett.109.266406}
\bibinfo{author}{\bibfnamefont{R.}~\bibnamefont{Comin}},
  \bibinfo{author}{\bibfnamefont{G.}~\bibnamefont{Levy}},
  \bibinfo{author}{\bibfnamefont{B.}~\bibnamefont{Ludbrook}},
  \bibinfo{author}{\bibfnamefont{Z.-H.} \bibnamefont{Zhu}},
  \bibinfo{author}{\bibfnamefont{C.}~\bibnamefont{Veenstra}},
  \bibinfo{author}{\bibfnamefont{J.}~\bibnamefont{Rosen}},
  \bibinfo{author}{\bibfnamefont{Y.}~\bibnamefont{Singh}},
  \bibinfo{author}{\bibfnamefont{P.}~\bibnamefont{Gegenwart}},
  \bibinfo{author}{\bibfnamefont{D.}~\bibnamefont{Stricker}},
  \bibinfo{author}{\bibfnamefont{J.}~\bibnamefont{Hancock}},
  \bibnamefont{et~al.}, \bibinfo{journal}{Phys. Rev. Lett.}
  \textbf{\bibinfo{volume}{109}}, \bibinfo{pages}{266406}
  (\bibinfo{year}{2012}).

\bibitem[{\citenamefont{Choi et~al.}(2012)\citenamefont{Choi, Coldea,
  Kolmogorov, Lancaster, Mazin, Blundell, Radaelli, Singh, Gegenwart, Choi
  et~al.}}]{PhysRevLett.108.127204}
\bibinfo{author}{\bibfnamefont{S.~K.} \bibnamefont{Choi}},
  \bibinfo{author}{\bibfnamefont{R.}~\bibnamefont{Coldea}},
  \bibinfo{author}{\bibfnamefont{A.~N.} \bibnamefont{Kolmogorov}},
  \bibinfo{author}{\bibfnamefont{T.}~\bibnamefont{Lancaster}},
  \bibinfo{author}{\bibfnamefont{I.~I.} \bibnamefont{Mazin}},
  \bibinfo{author}{\bibfnamefont{S.~J.} \bibnamefont{Blundell}},
  \bibinfo{author}{\bibfnamefont{P.~G.} \bibnamefont{Radaelli}},
  \bibinfo{author}{\bibfnamefont{Y.}~\bibnamefont{Singh}},
  \bibinfo{author}{\bibfnamefont{P.}~\bibnamefont{Gegenwart}},
  \bibinfo{author}{\bibfnamefont{K.~R.} \bibnamefont{Choi}},
  \bibnamefont{et~al.}, \bibinfo{journal}{Phys. Rev. Lett.}
  \textbf{\bibinfo{volume}{108}}, \bibinfo{pages}{127204}
  (\bibinfo{year}{2012}).

\bibitem[{\citenamefont{Ye et~al.}(2012)\citenamefont{Ye, Chi, Cao,
  Chakoumakos, Fernandez-Baca, Custelcean, Qi, Korneta, and
  Cao}}]{PhysRevB.85.180403}
\bibinfo{author}{\bibfnamefont{F.}~\bibnamefont{Ye}},
  \bibinfo{author}{\bibfnamefont{S.}~\bibnamefont{Chi}},
  \bibinfo{author}{\bibfnamefont{H.}~\bibnamefont{Cao}},
  \bibinfo{author}{\bibfnamefont{B.~C.} \bibnamefont{Chakoumakos}},
  \bibinfo{author}{\bibfnamefont{J.~A.} \bibnamefont{Fernandez-Baca}},
  \bibinfo{author}{\bibfnamefont{R.}~\bibnamefont{Custelcean}},
  \bibinfo{author}{\bibfnamefont{T.~F.} \bibnamefont{Qi}},
  \bibinfo{author}{\bibfnamefont{O.~B.} \bibnamefont{Korneta}},
  \bibnamefont{and} \bibinfo{author}{\bibfnamefont{G.}~\bibnamefont{Cao}},
  \bibinfo{journal}{Phys. Rev. B} \textbf{\bibinfo{volume}{85}},
  \bibinfo{pages}{180403} (\bibinfo{year}{2012}).

\bibitem[{\citenamefont{Singh et~al.}(2012)\citenamefont{Singh, Manni, Reuther,
  Berlijn, Thomale, Ku, Trebst, and Gegenwart}}]{PhysRevLett.108.127203}
\bibinfo{author}{\bibfnamefont{Y.}~\bibnamefont{Singh}},
  \bibinfo{author}{\bibfnamefont{S.}~\bibnamefont{Manni}},
  \bibinfo{author}{\bibfnamefont{J.}~\bibnamefont{Reuther}},
  \bibinfo{author}{\bibfnamefont{T.}~\bibnamefont{Berlijn}},
  \bibinfo{author}{\bibfnamefont{R.}~\bibnamefont{Thomale}},
  \bibinfo{author}{\bibfnamefont{W.}~\bibnamefont{Ku}},
  \bibinfo{author}{\bibfnamefont{S.}~\bibnamefont{Trebst}}, \bibnamefont{and}
  \bibinfo{author}{\bibfnamefont{P.}~\bibnamefont{Gegenwart}},
  \bibinfo{journal}{Phys. Rev. Lett.} \textbf{\bibinfo{volume}{108}},
  \bibinfo{pages}{127203} (\bibinfo{year}{2012}).

\bibitem[{\citenamefont{Reuther et~al.}(2011)\citenamefont{Reuther, Thomale,
  and Trebst}}]{PhysRevB.84.100406}
\bibinfo{author}{\bibfnamefont{J.}~\bibnamefont{Reuther}},
  \bibinfo{author}{\bibfnamefont{R.}~\bibnamefont{Thomale}}, \bibnamefont{and}
  \bibinfo{author}{\bibfnamefont{S.}~\bibnamefont{Trebst}},
  \bibinfo{journal}{Phys. Rev. B} \textbf{\bibinfo{volume}{84}},
  \bibinfo{pages}{100406} (\bibinfo{year}{2011}).

\bibitem[{\citenamefont{Chaloupka et~al.}(2013)\citenamefont{Chaloupka,
  Jackeli, and Khaliullin}}]{PhysRevLett.110.097204}
\bibinfo{author}{\bibfnamefont{J.}~\bibnamefont{Chaloupka}},
  \bibinfo{author}{\bibfnamefont{G.}~\bibnamefont{Jackeli}}, \bibnamefont{and}
  \bibinfo{author}{\bibfnamefont{G.}~\bibnamefont{Khaliullin}},
  \bibinfo{journal}{Phys. Rev. Lett.} \textbf{\bibinfo{volume}{110}},
  \bibinfo{pages}{097204} (\bibinfo{year}{2013}).

\bibitem[{\citenamefont{Kimchi and You}(2011)}]{PhysRevB.84.180407}
\bibinfo{author}{\bibfnamefont{I.}~\bibnamefont{Kimchi}} \bibnamefont{and}
  \bibinfo{author}{\bibfnamefont{Y.-Z.} \bibnamefont{You}},
  \bibinfo{journal}{Phys. Rev. B} \textbf{\bibinfo{volume}{84}},
  \bibinfo{pages}{180407} (\bibinfo{year}{2011}).

\bibitem[{\citenamefont{Albuquerque et~al.}(2011)\citenamefont{Albuquerque,
  Schwandt, Het\'enyi, Capponi, Mambrini, and L\"auchli}}]{PhysRevB.84.024406}
\bibinfo{author}{\bibfnamefont{A.~F.} \bibnamefont{Albuquerque}},
  \bibinfo{author}{\bibfnamefont{D.}~\bibnamefont{Schwandt}},
  \bibinfo{author}{\bibfnamefont{B.}~\bibnamefont{Het\'enyi}},
  \bibinfo{author}{\bibfnamefont{S.}~\bibnamefont{Capponi}},
  \bibinfo{author}{\bibfnamefont{M.}~\bibnamefont{Mambrini}}, \bibnamefont{and}
  \bibinfo{author}{\bibfnamefont{A.~M.} \bibnamefont{L\"auchli}},
  \bibinfo{journal}{Phys. Rev. B} \textbf{\bibinfo{volume}{84}},
  \bibinfo{pages}{024406} (\bibinfo{year}{2011}).

\bibitem[{\citenamefont{Bhattacharjee et~al.}(2012)\citenamefont{Bhattacharjee,
  Lee, and Kim}}]{bhattacharjee2012spin}
\bibinfo{author}{\bibfnamefont{S.}~\bibnamefont{Bhattacharjee}},
  \bibinfo{author}{\bibfnamefont{S.-S.} \bibnamefont{Lee}}, \bibnamefont{and}
  \bibinfo{author}{\bibfnamefont{Y.~B.} \bibnamefont{Kim}},
  \bibinfo{journal}{New Journal of Physics} \textbf{\bibinfo{volume}{14}},
  \bibinfo{pages}{073015} (\bibinfo{year}{2012}).

\bibitem[{\citenamefont{Mazin et~al.}(2012)\citenamefont{Mazin, Jeschke,
  Foyevtsova, Valent\'\i, and Khomskii}}]{PhysRevLett.109.197201}
\bibinfo{author}{\bibfnamefont{I.~I.} \bibnamefont{Mazin}},
  \bibinfo{author}{\bibfnamefont{H.~O.} \bibnamefont{Jeschke}},
  \bibinfo{author}{\bibfnamefont{K.}~\bibnamefont{Foyevtsova}},
  \bibinfo{author}{\bibfnamefont{R.}~\bibnamefont{Valent\'\i}},
  \bibnamefont{and} \bibinfo{author}{\bibfnamefont{D.~I.}
  \bibnamefont{Khomskii}}, \bibinfo{journal}{Phys. Rev. Lett.}
  \textbf{\bibinfo{volume}{109}}, \bibinfo{pages}{197201}
  (\bibinfo{year}{2012}).

\bibitem[{\citenamefont{Miyake and Imada}(2010)}]{MiyakeReview2010}
\bibinfo{author}{\bibfnamefont{T.}~\bibnamefont{Miyake}} \bibnamefont{and}
  \bibinfo{author}{\bibfnamefont{M.}~\bibnamefont{Imada}}, \bibinfo{journal}{J.
  Phys. Soc. Jpn.} \textbf{\bibinfo{volume}{79}}, \bibinfo{pages}{112001}
  (\bibinfo{year}{2010}).

\bibitem[{\citenamefont{http://elk.sorceforge.net/}()}]{Elk}
\bibinfo{author}{\bibnamefont{http://elk.sorceforge.net/}}.

\bibitem[{\citenamefont{Perdew and Wang}(1992)}]{PhysRevB.45.13244}
\bibinfo{author}{\bibfnamefont{J.~P.} \bibnamefont{Perdew}} \bibnamefont{and}
  \bibinfo{author}{\bibfnamefont{Y.}~\bibnamefont{Wang}},
  \bibinfo{journal}{Phys. Rev. B} \textbf{\bibinfo{volume}{45}},
  \bibinfo{pages}{13244} (\bibinfo{year}{1992}).

\bibitem[{\citenamefont{Arita et~al.}(2012)\citenamefont{Arita,
  Kune\ifmmode~\check{s}\else \v{s}\fi{}, Kozhevnikov, Eguiluz, and
  Imada}}]{PhysRevLett.108.086403}
\bibinfo{author}{\bibfnamefont{R.}~\bibnamefont{Arita}},
  \bibinfo{author}{\bibfnamefont{J.}~\bibnamefont{Kune\ifmmode~\check{s}\else
  \v{s}\fi{}}}, \bibinfo{author}{\bibfnamefont{A.~V.}
  \bibnamefont{Kozhevnikov}}, \bibinfo{author}{\bibfnamefont{A.~G.}
  \bibnamefont{Eguiluz}}, \bibnamefont{and}
  \bibinfo{author}{\bibfnamefont{M.}~\bibnamefont{Imada}},
  \bibinfo{journal}{Phys. Rev. Lett.} \textbf{\bibinfo{volume}{108}},
  \bibinfo{pages}{086403} (\bibinfo{year}{2012}).

\bibitem[{\citenamefont{Aryasetiawan et~al.}(2004)\citenamefont{Aryasetiawan,
  Imada, Georges, Gabriel, Biermann, and Lichtenstein}}]{Aryasetiawan}
\bibinfo{author}{\bibfnamefont{F.}~\bibnamefont{Aryasetiawan}},
  \bibinfo{author}{\bibfnamefont{M.}~\bibnamefont{Imada}},
  \bibinfo{author}{\bibfnamefont{A.}~\bibnamefont{Georges}},
  \bibinfo{author}{\bibfnamefont{K.}~\bibnamefont{Gabriel}},
  \bibinfo{author}{\bibfnamefont{S.}~\bibnamefont{Biermann}}, \bibnamefont{and}
  \bibinfo{author}{\bibfnamefont{A.~I.} \bibnamefont{Lichtenstein}},
  \bibinfo{journal}{Phys. Rev. B} \textbf{\bibinfo{volume}{70}},
  \bibinfo{pages}{195104} (\bibinfo{year}{2004}).

\bibitem[{\citenamefont{Kozhevnikov et~al.}(2010)\citenamefont{Kozhevnikov,
  Eguiluz, and Schulthess}}]{AntoncRPA}
\bibinfo{author}{\bibfnamefont{A.}~\bibnamefont{Kozhevnikov}},
  \bibinfo{author}{\bibfnamefont{A.}~\bibnamefont{Eguiluz}}, \bibnamefont{and}
  \bibinfo{author}{\bibfnamefont{T.}~\bibnamefont{Schulthess}},
  \bibinfo{journal}{SC'10 Proceedings of the 2010 ACM/IEEE International
  Conference for High Performance Computing, Networking, Storage, and Analysis}
  p.~\bibinfo{pages}{1} (\bibinfo{year}{2010}).

\bibitem[{\citenamefont{Katukuri et~al.}(2014)\citenamefont{Katukuri,
  Nishimoto, Yushankhai, Stoyanova, Kandpal, Sungkyun, Coldea, Rousochatzakis,
  Hozoi, and van~den Brink}}]{arXiv:1312.7437}
\bibinfo{author}{\bibfnamefont{V.~K.} \bibnamefont{Katukuri}},
  \bibinfo{author}{\bibfnamefont{S.}~\bibnamefont{Nishimoto}},
  \bibinfo{author}{\bibfnamefont{V.}~\bibnamefont{Yushankhai}},
  \bibinfo{author}{\bibfnamefont{A.}~\bibnamefont{Stoyanova}},
  \bibinfo{author}{\bibfnamefont{H.}~\bibnamefont{Kandpal}},
  \bibinfo{author}{\bibfnamefont{C.}~\bibnamefont{Sungkyun}},
  \bibinfo{author}{\bibfnamefont{R.}~\bibnamefont{Coldea}},
  \bibinfo{author}{\bibfnamefont{I.}~\bibnamefont{Rousochatzakis}},
  \bibinfo{author}{\bibfnamefont{L.}~\bibnamefont{Hozoi}}, \bibnamefont{and}
  \bibinfo{author}{\bibfnamefont{J.}~\bibnamefont{van~den Brink}},
  \bibinfo{journal}{New J. Phys.} \textbf{\bibinfo{volume}{16}},
  \bibinfo{pages}{013056} (\bibinfo{year}{2014}).

\bibitem[{Not()}]{Note1}
\bibinfo{note}{The small amplitude of the trigonal distortion $\Delta=-28$ meV
  is different from the previous result\cite{PhysRevLett.108.106401}. The
  difference originates from the electronic band structures around the Fermi
  level in Ref.\onlinecite{PhysRevLett.108.106401} different from the result in
  Ref.15 and ours. In addition, in Ref.\onlinecite{PhysRevLett.108.106401}, the
  tight-binding parameter is obtained through fitting the LDA dispersion, while
  we directly calculate the hopping matrix elements with the $t_{2g}$ Wannier
  orbitals.}

\bibitem[{\citenamefont{Sugiura and Shimizu}(2012)}]{PhysRevLett.108.240401}
\bibinfo{author}{\bibfnamefont{S.}~\bibnamefont{Sugiura}} \bibnamefont{and}
  \bibinfo{author}{\bibfnamefont{A.}~\bibnamefont{Shimizu}},
  \bibinfo{journal}{Phys. Rev. Lett.} \textbf{\bibinfo{volume}{108}},
  \bibinfo{pages}{240401} (\bibinfo{year}{2012}).

\bibitem[{\citenamefont{Jakli\ifmmode~\check{c}\else \v{c}\fi{} and
  Prelov\ifmmode~\check{s}\else \v{s}\fi{}ek}(1994)}]{PhysRevB.49.5065}
\bibinfo{author}{\bibfnamefont{J.}~\bibnamefont{Jakli\ifmmode~\check{c}\else
  \v{c}\fi{}}} \bibnamefont{and}
  \bibinfo{author}{\bibfnamefont{P.}~\bibnamefont{Prelov\ifmmode~\check{s}\else
  \v{s}\fi{}ek}}, \bibinfo{journal}{Phys. Rev. B}
  \textbf{\bibinfo{volume}{49}}, \bibinfo{pages}{5065} (\bibinfo{year}{1994}).

\bibitem[{\citenamefont{Imada and Takahashi}(1986)}]{Imada_Takahashi}
\bibinfo{author}{\bibfnamefont{M.}~\bibnamefont{Imada}} \bibnamefont{and}
  \bibinfo{author}{\bibfnamefont{M.}~\bibnamefont{Takahashi}},
  \bibinfo{journal}{J. Phys. Soc. Jpn.} \textbf{\bibinfo{volume}{55}},
  \bibinfo{pages}{3354} (\bibinfo{year}{1986}).

\bibitem[{\citenamefont{Rau et~al.}(2014)\citenamefont{Rau, Lee, and
  Kee}}]{arXiv:1310.7940}
\bibinfo{author}{\bibfnamefont{J.~G.} \bibnamefont{Rau}},
  \bibinfo{author}{\bibfnamefont{E.~K.-H.} \bibnamefont{Lee}},
  \bibnamefont{and} \bibinfo{author}{\bibfnamefont{H.-Y.} \bibnamefont{Kee}},
  \bibinfo{journal}{Phys. Rev. Lett.} \textbf{\bibinfo{volume}{112}},
  \bibinfo{pages}{077204} (\bibinfo{year}{2014}).

\bibitem[{\citenamefont{Cao et~al.}(2013)\citenamefont{Cao, Qi, Li, Terzic,
  Cao, Yuan, Tovar, Murthy, and Kaul}}]{PhysRevB.88.220414}
\bibinfo{author}{\bibfnamefont{G.}~\bibnamefont{Cao}},
  \bibinfo{author}{\bibfnamefont{T.~F.} \bibnamefont{Qi}},
  \bibinfo{author}{\bibfnamefont{L.}~\bibnamefont{Li}},
  \bibinfo{author}{\bibfnamefont{J.}~\bibnamefont{Terzic}},
  \bibinfo{author}{\bibfnamefont{V.~S.} \bibnamefont{Cao}},
  \bibinfo{author}{\bibfnamefont{S.~J.} \bibnamefont{Yuan}},
  \bibinfo{author}{\bibfnamefont{M.}~\bibnamefont{Tovar}},
  \bibinfo{author}{\bibfnamefont{G.}~\bibnamefont{Murthy}}, \bibnamefont{and}
  \bibinfo{author}{\bibfnamefont{R.~K.} \bibnamefont{Kaul}},
  \bibinfo{journal}{Phys. Rev. B} \textbf{\bibinfo{volume}{88}},
  \bibinfo{pages}{220414} (\bibinfo{year}{2013}).

\bibitem[{\citenamefont{Singh and Gegenwart}(2010)}]{Singh_Gegenwart}
\bibinfo{author}{\bibfnamefont{Y.}~\bibnamefont{Singh}} \bibnamefont{and}
  \bibinfo{author}{\bibfnamefont{P.}~\bibnamefont{Gegenwart}},
  \bibinfo{journal}{Phys. Rev. B} \textbf{\bibinfo{volume}{82}},
  \bibinfo{pages}{064412} (\bibinfo{year}{2010}).

\bibitem[{\citenamefont{Kim et~al.}(2012)\citenamefont{Kim, Kim, Jeong, Jin,
  and Yu}}]{PhysRevLett.108.106401}
\bibinfo{author}{\bibfnamefont{C.~H.} \bibnamefont{Kim}},
  \bibinfo{author}{\bibfnamefont{H.~S.} \bibnamefont{Kim}},
  \bibinfo{author}{\bibfnamefont{H.}~\bibnamefont{Jeong}},
  \bibinfo{author}{\bibfnamefont{H.}~\bibnamefont{Jin}}, \bibnamefont{and}
  \bibinfo{author}{\bibfnamefont{J.}~\bibnamefont{Yu}}, \bibinfo{journal}{Phys.
  Rev. Lett.} \textbf{\bibinfo{volume}{108}}, \bibinfo{pages}{106401}
  (\bibinfo{year}{2012}).

\end{thebibliography}
\end{document}